\documentclass[pre,twocolumn,aps,amsmath,floatfix,showpacs,amsfonts,amssymb,10pt]{revtex4-1}
\usepackage{graphicx}
\usepackage{color}
\usepackage{times}
\usepackage[utf8]{inputenc}
\usepackage[english]{babel}
\usepackage{fancyhdr}
\usepackage{marvosym}
\usepackage{hyperref}
\usepackage{amsmath}
\usepackage{mathrsfs}
\usepackage{latexsym}
\usepackage{amssymb}
\usepackage{graphicx} 
\newcommand{\Det}[1]{\det \left[ #1 \right]}
\newcommand{\Tr}[1]{\text{Tr}\left[ #1\right]}
\newcommand{\kT}{k_{B}T}
\newcommand{\hc}{\hbar c}
\newcommand{\PFA}{\mathrm{PFA}}

\begin{document}

\title{Effect of Curvature and Confinement on the Casimir-Polder Interaction}

\author{Pablo Rodriguez-Lopez}
\affiliation{Laboratoire de Physique Th\'eorique et Mod\`eles
  Statistiques, CNRS UMR 8626, B\^at.~100, Universit\'e Paris-Sud, 91405
  Orsay cedex, France}
  
\author{Thorsten Emig}
\affiliation{Laboratoire de Physique Th\'eorique et Mod\`eles
  Statistiques, CNRS UMR 8626, B\^at.~100, Universit\'e Paris-Sud, 91405
  Orsay cedex, France}

\author{Ehsan Noruzifar}
\affiliation{Department of Physics and Astronomy, University of California, Riverside, California 92521, USA}

\author{Roya Zandi}
\affiliation{Department of Physics and Astronomy, University of California, Riverside, California 92521, USA}

\begin{abstract} 
Modifications of Casimir-Polder interactions due to confinement inside a cylindrical cavity and due to curvature in- and outside the cavity are studied.  We consider a perfectly conducting cylindrical shell with a single particle (atom or macroscopic sphere) located next to its interior or exterior surface, or two atoms placed inside the shell. By employing the scattering approach, we obtain the particle-cavity interaction and the modification of the two-particle interaction due to the cavity.  We consider both retardation and thermal effects.  While for the atoms a dipole description is sufficient, for the macroscopic sphere we sum (numerically) over many multipole fluctuations to compute the interaction at short separations. In the latter limit we compare to the proximity approximation and a gradient expansion and find agreement. Our results indicate an confinement induced suppression of the force between atoms. General criteria for suppression and enhancement of Casimir interactions due to confinement are 
discussed. 
\end{abstract}


\maketitle


\section{Introduction}

Fluctuation-induced interactions between polarizable particles such as the London force \cite{London} and its retarded companion, the Casimir-Polder force \cite{VdW int. electrica}, play an ubiquitous role in fundamental physics, chemistry and applied technologies including nanotechnology and ultra-cold atomic systems \cite{Serry,Klimchitskaya2009,Casimir_review}. A quantitative understanding of such forces is a key parameter in the design of new experiments and technological applications. 

An important line of research addresses the dependence of these interactions on geometry and shape of the polarizable objects. Examples include the effect of surface curvature \cite{Derivative_Expansion_PFA_energy}, sharp edges and tips \cite{edges_and_tips}, orientation dependence \cite{Orientation_dependence}, and confinement \cite{Object_in_spherical_cavity, cylinder_in_cylindrical_cavity}. The non-additivity of fluctuation induced forces complicates the study of these effects. To overcome this problem, a multiscattering formalism for Casimir forces has been developed \cite{EGJK,Rahi-Emig,Multiscattering_Lambrecht} and applied successfully \cite{Real_metallic_cylinders,Real_Cylinder_and_plane,Cylinder_and_plane,Cylinders_and_Plates,spheres_and_plates,inclined_cylinders}, including the case of a sphere outside a cylinder \cite{Ehsan,Teo_Sphere_Cylinder}.

Here we are interested mainly in confined geometries where the particles (atoms or a metallic sphere) are placed inside the a perfectly conducting cylindrical cavity. In such situation the force on an object at short distance from the confining walls is modified by curvature and the force between two particles at large distances (compared to the confinement scale) is different from the force in free space. Boundary effects on the force between a pair of atoms at zero temperature have been studied originally by Mahanty and Ninham \cite{2_atoms_in_PM_capacitor_zeroT}, showing a reduced force in the non-retarded limit and an increased force in the retarded limit. At any finite temperature, an exponentially reduced force was found
\cite{2_atoms_in_PM_capacitor_highT}. More recently, there have been studies of the Casimir-Polder interaction for an atom inside a cylindrical cavity \cite{Ellingsen1,PFA_Atom_Cylinder}, and a compact object inside a spherical cavity \cite{Object_in_spherical_cavity}. Recent studies of the modification of the force between two atoms inside quasi one-dimensional structures have displayed an exponential reduction of the interaction for a rectangular waveguide \cite{Atoms_in_rectangular_cavity} and a huge amplification of the interaction for two concentric metallic cylinders\cite{Shahmoon+2014}.

Here we show that the scattering approach can be used to obtain both a large and a short distance expansion for the force  between an atom that is placed either inside or outside a cylindrical cavity. Our results apply to the  zero temperature and the classical high temperature limit. Going beyond the small particle limit, we consider the interaction of a macroscopic perfect metal sphere with the wall of a confining cylindrical shell. The comparison of our numerical results with a gradient expansion of the interaction demonstrates nice agreement at small separations. For the interaction between two atoms that are placed on the axis of a confining cylindrical shell we derive exact results in the asymptotic limit of large separations both at zero and high temperatures. The observed exponential decay is explained in terms of evanescent modes and related to existing prediction in the literature.

The remainder of the article is organized as follows: In Sec.~\ref{Multiscattering formalism of Casimir energy} we briefly review the multiscattering approach used to obtain Casimir interactions. In Sec.~\ref{Casimir energy of an atom inside a metallic Cylinder} we analyze the retarded and non-retarded Casimir interaction between an atom and a perfect metal cylindrical cavity, both in the high and low temperature limits. Interior and exterior cases are considered. In Sec.~\ref{PFA of the Casimir energy of a sphere inside a metal Cylinder} the proximity force approximation (PFA) for a perfect metal sphere inside a perfect metal cylindrical cavity is given.  In Sec.~\ref{Casimir energy of a sphere inside a metal Cylinder - Numerical study}, the Casimir interaction for the latter geometry is computed numerically and compared to the PFA and a gradient expansion.  In Sec.~\ref{Effect of confinement and non--pairwise behavior of Casimir energy}, the confinement of two atoms is studied by computing their 
interaction along the axis of a perfect metal cylindrical cavity.  We conclude with a discussion of our results, and a summary of relevant matrix elements is given in App.~\ref{Appendix}.

\section{Scattering approach}
\label{Multiscattering formalism of Casimir energy}

We employ the scattering approach \cite{EGJK,Rahi-Emig} to compute the Casimir interactions. This approach relates the interaction between objects to their electromagnetic scattering properties. The Casimir free energy at temperature $T$ is given by
\begin{equation}
\label{Energy_T_finite}
E = \kT{\sum_{n=0}^{\infty}}'\log\det [ \mathcal{I} - \mathcal{N}(\kappa_{n})],
\end{equation}
where $\kappa_{n} = 2\pi n k_{B}T/\hc$ are Matsubara wave numbers. The prime indicates that the zero Matsubara frequency contribution has a weight of $1/2$. At zero temperature, the primed sum is replaced by an integral along the imaginary frequency axis, yielding the Casimir energy
\begin{equation}\label{zero T Casimir energy}
E_{0} = \frac{\hc}{2\pi}\int_{0}^{\infty}d\kappa \log\det [ \mathcal{I} - \mathcal{N}(\kappa)].
\end{equation}
The high temperature or classical limit is reached when the distance between the interacting objects becomes much larger than the thermal wavelength $\lambda_{T} = \hbar c/(k_{B}T)$. The Casimir free energy is then given by the zero Matsubara frequency term of Eq.~\eqref{Energy_T_finite},
\begin{equation}\label{high T Casimir energy}
E_{cl} = \frac{\kT}{2}\log\det [ \mathcal{I} - \mathcal{N}(0) ] \, .
\end{equation}
Following the notation of~\cite{Rahi-Emig,Object_in_spherical_cavity}, the blocks forming the matrix $\mathcal{N}$ for a geometry consisting of a finite number of objects are given by
\begin{equation}
\left(\mathcal{I} - \mathcal{N}\right)_{\alpha\beta} = \delta_{\alpha\beta} + (1 - \delta_{\alpha\beta})\mathcal{T}_{\alpha}\mathcal{X}_{\alpha\beta},
\end{equation}
where $\alpha$ and $\beta$ label the interacting bodies. $\mathcal{T}_{\alpha}$ is the T-matrix of object $\alpha$ (either the cavity or an object inside), and it encodes all information about shape and material composition of the object. We are interested in a situation where objects are entirely enclosed within the cavity formed by an external object which will be a cylindrical shell. The T-matrix of the external object relevant to this situation of internal scattering is different from the T-matrix of the same object for external scattering. The latter case is relevant to objects placed outside the cylinder. For perfectly conducting boundary conditions, the T-matrix for internal scattering is the inverse of the T-matrix for external scattering. The translation matrices $\mathcal{X}_{\alpha\beta}$ describe the interaction of the fluctuating multipole moments on object $\alpha$ and $\beta$. They contain all information about the relative position and orientation of the objects. If $\mathcal{X}_{\alpha\beta}$ connects two separate objects (both either inside or outside of a cavity), we have $\mathcal{X}_{\alpha\beta} = \mathcal{U}_{\alpha\beta}$, where the matrix $\mathcal{U}_{\alpha\beta}$ relates regular waves to outgoing waves \cite{Rahi-Emig}. When $\mathcal{X}_{\alpha\gamma}$ connects a body $\alpha$ to the external cavity $\gamma = C$ ($C$ will denote the cavity in the following) enclosing it, we have $\mathcal{X}_{\alpha C} = \mathcal{V}_{\alpha C}$ where the matrix $\mathcal{V}_{\alpha C}$ connects regular waves (with respect to the origin of coordinates) of the object $\alpha$ to regular waves of the cavity $C$ \cite{Rahi-Emig}. Finally, if $\mathcal{X}_{C \beta}$ connects the enclosing cavity $C$ with a body $\beta$ placed in its interior, we have $\mathcal{X}_{C\beta} = \mathcal{W}_{C\beta}$, relating now regular waves of the cavity $C$ to regular waves of $\beta$, with $\mathcal{W}^{ij}_{C\beta} =\mathcal{V}^{\dagger\,ij}_{\beta C}\frac{C_i}{C_j}$ where the generalized indices $i$, $j$ label the multipole waves including polarization, and $C_i$ are normalization constants \cite{Rahi-Emig}.
For the configuration of one object $S$ inside or outside the cavity $C$, the  matrix $\mathcal{N}$ is given by
\begin{equation}\label{N_Matrix_1object_Internal_to_a_Cavity}
\mathcal{N} = \mathcal{T}_S\mathcal{X}_{SC}\mathcal{T}_C\mathcal{X}_{CS} \, .
\end{equation}
The T-matrix of object $S$ and the cavity $C$ are usually given in different vector multipole basis.  Therefore, we have to apply a change of basis to one of the T-matrices in order to apply the above formula. Since here the cavity is a cylindrical shell and the particle $S$ has spherical symmetry (atom or sphere), 
we shall use the conversion matrices $D_{l,m,P,n,k_{z},Q}$ of spherical vector waves $\phi_{l, m,P}$ into cylindrical vector waves $\phi_{n,k_{z},Q} $ defined by \cite{Ehsan}
\begin{equation}\label{TMM_general}
\phi_{l, m,P} = \sum_{n\in\mathbb{Z}}\sum_{Q=M,N}\int_{-\infty}^{\infty}\frac{dk_{z}}{2\pi}D_{l,m,P,n,k_{z},Q}\, \phi_{n,k_{z},Q}\ \, ,
\end{equation}
where $Q$ denotes electric ($N$) or magnetic ($M$) polarizations.
By the use of these conversion matrices, the T-matrix of object $S$ (given in spherical wave basis) can be transformed to the cylindrical vector basis as~\cite{Rahi-Emig}
\begin{eqnarray}
\mathcal{T}_{S\,\, n\,k_{z}\,P,n'\,k'_{z}\,P'} & = & \sum_{l, m, Q}\sum_{l', m',Q'}
\frac{C^{c}_{P}}{C^{s}_{Q}}
D^{\dagger}_{n\,k_{z}\,P,l m Q} \nonumber\\
& &\times\mathcal{T}_{S\,\,l\,m\,Q,l'\,m'\,Q'}
D_{l'\,m'\,Q',n'\,k'_{z}\,P'},\label{Sph_2_Cyl}
\end{eqnarray}
where we have used $C^{c}_{M} = -\frac{1}{2\pi} = - C^{c}_{N}$ and $C^{s}_{M} = \kappa = - C^{s}_{N}$ to obtain $C^{c}_{P}/C^{s}_{Q} = (2\pi\kappa)^{-1}\left( 1 - 2\delta_{P,Q} \right)$ \cite{Rahi-Emig} (see also App.~\ref{Appendix}). Now the matrix $\mathcal{N}$ for a compact object inside a cylindrical cavity can be expressed in spherical vector wave basis as
\begin{eqnarray}\label{Nmatrix_sphere_in_cylindrical_cavity}
&& \mathcal{N}_{l'm',l m} =  \frac{1}{4\pi^{2}\kappa}\mathcal{T}_{S\,\,l'm',lm}\int_{-\infty}^{\infty}dk_{z}D_{lm,nk_{z}}\\
&  & \times \mathcal{V}_{SC\,\,nk_{z},n'k'_{z}}\mathcal{T}_{C\,\,n'k'_{z},n''k''_{z}}\mathcal{V}^{\dagger}_{CS \,\,n''k''_{z},\tilde{n}\tilde{k}_{z}}D^{\dagger}_{\tilde{n}\tilde{k}_{z},l m},\nonumber
\end{eqnarray}
with summation over repeated indices.
Note that the polarization indices are not explicitly shown. Details about the matrices used in Eq.~\eqref{Nmatrix_sphere_in_cylindrical_cavity} can be found in App.~\ref{Appendix}. The energies defined by Eqs.~\eqref{Energy_T_finite}, \eqref{zero T Casimir energy} and \eqref{high T Casimir energy} are finite for all positions of the internal object inside the cylinder. Therefore, a regularization is not needed, in contrast to the interaction of two exterior objects where the energy for infinite distance is usually subtracted.

\section{Casimir energy of an atom and a metallic Cylinder}
\label{Casimir energy of an atom inside a metallic Cylinder}

In this section we calculate the Casimir energy of an atom that is placed outside or inside a perfect metal cylindrical cavity.  We obtain analytical results in asymptotic regimes, and numerical results for intermediate distances.  Our results are based on two assumptions: (1) The atom is described by its electric and magnetic dipolar polarizabilities $\alpha_E$ and $\alpha_M$, respectively, and has no higher multipole polarizabilities, and (2) the polarizabilities are small compared to all geometric length scales, i.e., $\alpha_P^{1/3} \ll d, R, \ell$.  These assumptions are frequently made in the literature, leading to the Casimir-Polder interaction between a flat surface and an atom \cite{VdW int. electrica}. The second assumptions justifies to consider only one scattering at the atom so that the energy is linear in $\alpha_P$, and hence we keep only the linear term in
\begin{equation}
\label{Taylor_Series_on_small_N_in}
\log\Det{\mathcal{I} - \mathcal{N}} = - \Tr{\mathcal{N}} +{\cal O}(\mathcal{N}^2) \, .
\end{equation}
The first assumption implies that the trace runs only over dipolar waves $(l=1)$, yielding
\begin{eqnarray}\label{N_Matrix_1_atoms_inside_a_cylindrical_cavity_zeroT_indices}
\Tr{{\mathcal{N}}} & =& \frac{1}{4\pi^{2}\kappa}\mathcal{T}_{S\,1,m,P;1, m',P'} D_{1,m',P';n,k_{z},P''}\nonumber\\
& & \times \mathcal{X}_{SC,\,\,n,k_{z},P'';n',k'_{z},Q}\mathcal{T}_{C\,\,n',k'_{z},Q;n'',k''_{z},Q'}\nonumber\\
& & \times \mathcal{X}_{CS,\,\,n'',k''_{z},Q';\tilde{n},\tilde{k}_{z},Q''}D^{\dagger}_{\tilde{n},\tilde{k}_{z},Q'';1,m,P},
\end{eqnarray}
where summation over all discrete indices and integration over all wave vectors along the $z$-axis is assumed.
Here $\mathcal{T}_S$ represents the atom (for the T-matrix of an atom, see App.~\ref{Appendix}.)
In the following we shall apply these general results to a perfectly conducting cylindrical cavity.

\subsection{Retarded Casimir-Polder energy of an atom inside a cylinder}

\subsubsection{Zero temperature limit}

The Casimir energy at zero temperature of the atom placed at a distance $d<R_c$ from the axis of the cylindrical shell of radius $R_c$ can be written as 
\begin{equation}
E_{0} = - \frac{\hc}{R_{c}^{4}}\left(f_{0}^{E}(\delta)\alpha_{E} + f_{0}^{M}(\delta)\alpha_{M}\right),
\end{equation}
where $\delta = d/R_{c}$. The dependence on $\delta$ is determined by the following auxiliary functions
\begin{equation}\label{AuxI}
  \mathcal{I}_{\alpha,m}^{P}(\delta) = \int_{0}^{\infty}dt\,t^{\alpha}\sum_{n\in\mathbb{Z}}
 I_{n+m}^{2}(t\delta)\mathcal{T}_{C\,\,n,P;n,P}(t),
\end{equation}
with the Bessel function $I_n$ and where $\mathcal{T}_C$ is the T-matrix that describes the scattering inside the perfectly conducting cylindrical shell where the integration variable $t$ stands for the rescaled expression $\sqrt{\kappa^2 + k_z^2}$. 
The diagonal matrix elements are given by (see App.~\ref{Appendix})
\begin{align}
\mathcal{T}_{C\,\,n,M;n,M}(t) &= - \frac{K'_{n}(t)}{I'_{n}(t)} \, ,\\
\mathcal{T}_{C \,\,n,N;n,N}(t) &= - \frac{K_{n}(t)}{I_{n}(t)} \, .
\end{align}
The amplitudes $f_{0}^{E}(\delta)$ and $f_{0}^{M}(\delta)$ are then given by
\begin{align}
\label{fE0}
f_{0}^{E}(\delta) & = \frac{1}{4\pi}\left( \mathcal{I}_{3,1}^{M}(\delta) - \mathcal{I}_{3,1}^{N}(\delta) - 2 \mathcal{I}_{3,0}^{N}(\delta) \right)\, ,\\
\label{fM0}
f_{0}^{M}(\delta) & = - \frac{1}{4\pi}\left( \mathcal{I}_{3,1}^{M}(\delta) - \mathcal{I}_{3,1}^{N}(\delta) + 2 \mathcal{I}_{3,0}^{M}(\delta) \right)\, .
\end{align}
Using the uniform asymptotic expansion for Bessel functions as employed in \cite{PFA_Atom_Cylinder}, we 
obtain for {\it small} atom-surface distance $\ell=R_c-d$ the expansion
\begin{align}
\label{short_distance_limit_fE0}
f_{0}^{E} & = \frac{3}{8\pi} \left(\frac{R_c}{\ell}\right)^4 + \frac{13}{60\pi} \left(\frac{R_c}{\ell}\right)^3 + \frac{311}{1680\pi}\left(\frac{R_c}{\ell}\right)^2 + \cdots \, ,\\
\label{short_distance_limit_fM0}
f_{0}^{M}  & = -\frac{3}{8\pi} \left(\frac{R_c}{\ell}\right)^4 - \frac{17}{60\pi} \left(\frac{R_c}{\ell}\right)^3 - \frac{431}{1680\pi}  \left(\frac{R_c}{\ell}\right)^2 + \cdots \, .
\end{align}
The leading term agrees with the Casimir-Polder interaction between an atom and a planar surface \cite{VdW int. electrica}. The sub-leading terms describe curvature corrections. The first correction term for the $E$ polarization 
has been obtained in Ref.~\cite{PFA_Atom_Cylinder}.

When the atom is close to the center of the cylinder ($\delta\ll 1$), one can expand the functions $\mathcal{I}_{\alpha,m}^{P}(\delta)$ for small $\delta$ and perform the integral over $t$ and infinite sum over $n$ numerically. The result is 
\begin{align}
\label{fE0delta0}
f_{0}^{E}(\delta) &= \phantom{-} 0.594032 + 3.67884\,\delta^{2} +{\cal O}(\delta^4)\, , \\
\label{fM0delta0}
f_{0}^{M}(\delta) &=  - 0.805032 - 4.15129\,\delta^{2}+{\cal O}(\delta^4) \, .
\end{align}
Note that for atoms without magnetic response ($\alpha_{M} = 0$ and $\alpha_{E} > 0$), the Casimir-Polder interaction is attractive, while for (fictive) atoms without electric response ($\alpha_{E}=0$ and $\alpha_{M}>0$), the interaction is repulsive. This is in agreement with the interaction between an atom and a flat surface. Hence, surface curvature does not change the sign of the interaction. 

\begin{figure}
\begin{center}
\includegraphics[width=1.0\columnwidth]{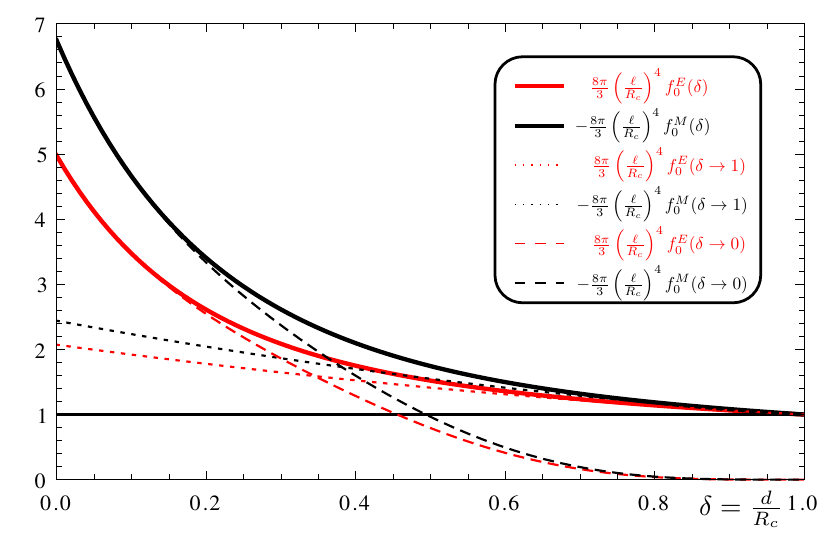} \caption{(Color online). Amplitudes of the interior Casimir-Polder interaction between an atom and a cylindrical shell as function of $\delta = d/R_{c}$. Shown are the limiting analytical results for short distances (dotted curves, see Eqs.~\eqref{short_distance_limit_fE0}, \eqref{short_distance_limit_fM0}), and when the atom is close to the axis of the cylinder (dashed curves, see Eqs.~\eqref{fE0delta0}, \eqref{fM0delta0}). The thick curves correspond to the full numerical result for $f_{0}^{E}(\delta)$ (Eq.~\eqref{fE0}) and $f_{0}^{M}(\delta)$ (Eq.~\eqref{fM0}).}
\label{zeroT_Retarded_Casimir_energy_atom_inside_PM_Cylinder}
\end{center}
\end{figure}

The analytical results for the atom placed either close to the cylindrical surface or close to its axis are shown in Fig.~\ref{zeroT_Retarded_Casimir_energy_atom_inside_PM_Cylinder} together with numerical results valid for all distances $\delta$. This comparison shows that the two limiting results provide an accurate estimate of the interaction over a wide range of separations.

\subsubsection{High temperature limit}

When $d\gg \lambda_{T}$, the Casimir energy can be approximated by its high temperature limit, given by Eq.~\eqref{high T Casimir energy}. For an atom inside a cylindrical cavity, this expressions reduces to
\begin{equation}
E_{cl} = - \frac{\kT}{2}\Tr{{\mathcal{N}}(0)}\, ,
\end{equation}
which can be written as
\begin{equation}\label{E_Casimir_atomo_interno_a_cilindro_high_T}
E_{cl} = - \frac{\kT}{R_{c}^{3}}\left( f_{cl}^{E}(\delta)\alpha_{E} + f_{cl}^{M}(\delta)\alpha_{M} \right),
\end{equation}
The amplitudes $f_{cl}^{E}(\delta)$ and $f_{cl}^{M}(\delta)$ can be expressed in terms of $\mathcal{I}_{\alpha,m}^{P}(\delta)$ defined in Eq.~\eqref{AuxI} as
\begin{align}
\label{fEcl}
f_{cl}^{E}(\delta) & = - \frac{1}{\pi}\left( \mathcal{I}_{2,1}^{N}(\delta) + \mathcal{I}_{2,0}^{N}(\delta) \right)\, ,\\
\label{fMcl}
f_{cl}^{M}(\delta) & = - \frac{1}{\pi}\left( \mathcal{I}_{2,1}^{M}(\delta) + \mathcal{I}_{2,0}^{M}(\delta) \right)\, .
\end{align}
Using again the uniform asymptotic expansion for Bessel functions, we obtain the asymptotic behavior of $\mathcal{I}_{\alpha,m}^{P}(\delta)$ when the atom is close to the surface of the cavity. This yields the amplitudes for $\ell \ll R_c$,
\begin{align}
\label{short_distance_limit_fEcl}
f_{cl}^{E} &= \frac{1}{4} \left(\frac{R_c}{\ell}\right)^3 + \frac{1}{8} \left(\frac{R_c}{\ell}\right)^2 + 
\frac{1}{8} \left(\frac{R_c}{\ell}\right) + \cdots \, ,\\
\label{short_distance_limit_fMcl}
f_{cl}^{M} &= - \frac{1}{4} \left(\frac{R_c}{\ell}\right)^3 - \frac{1}{4} \left(\frac{R_c}{\ell}\right)^2 
 - \frac{1}{4} \left(\frac{R_c}{\ell}\right) + \cdots \, .
\end{align}
The leading terms describe the atom-planar surface interaction, and the additional terms are curvature corrections.

When the atom is close to the axis of the cylindrical cavity, a small $\delta$ expansion and subsequent 
numerical summation and integration yields the amplitudes
\begin{align}
\label{fEcldelta0}
f_{cl}^{E}(\delta) & = \phantom{-} 1.00274 + 4.00376\,\delta^{2} +{\cal O}(\delta^4)\, , \\
\label{fMcldelta0}
f_{cl}^{M}(\delta) & = - 1.79579 - 5.06599\,\delta^{2}+{\cal O}(\delta^4) \, .
\end{align}

\begin{figure}[h]
\begin{center}
\includegraphics[scale=1]{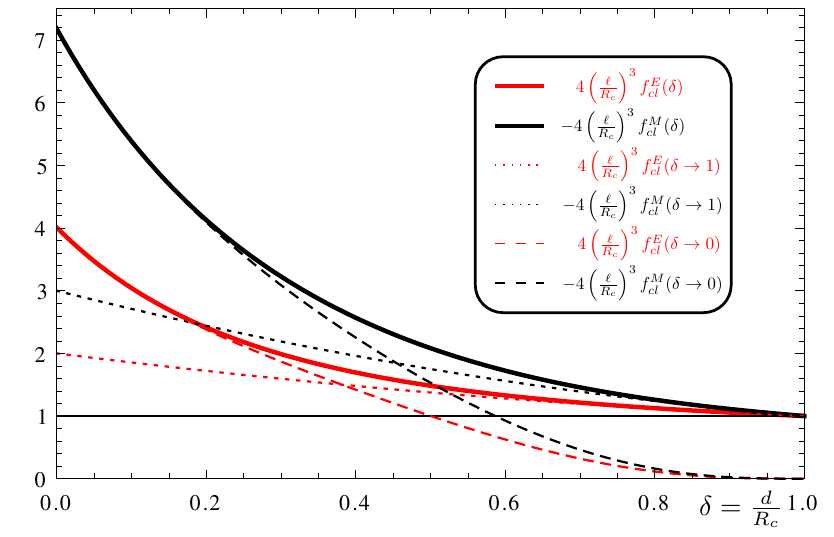}
\caption{(Color online). Equivalent of Fig.~\ref{zeroT_Retarded_Casimir_energy_atom_inside_PM_Cylinder}
for the classical limit. Shown are the analytical results for short surface distances $\ell$ (dotted curves, see Eqs.~\eqref{short_distance_limit_fEcl}, \eqref{short_distance_limit_fMcl}), and when the atom is close to the axis of the cylinder (dashed curves, see Eqs.~\eqref{fEcldelta0}, \eqref{fMcldelta0}). The thick curves represent the full numerical result for 
$f_{cl}^{E}(\delta)$ (Eq.~\eqref{fEcl}) and $f_{cl}^{M}(\delta)$ (Eq.~\eqref{fMcl}).}
\label{highT_Retarded_Casimir_energy_atom_inside_PM_Cylinder}
\end{center}
\end{figure}

We have also evaluated numerically the amplitudes of Eqs.~\eqref{fEcl} and \eqref{fMcl}. The results, and the approximations for small $\ell$ and small $\delta$ are shown in Fig.~\ref{highT_Retarded_Casimir_energy_atom_inside_PM_Cylinder}.

\subsection{Non-retarded London energy of an atom inside the cylinder}
\label{sec:London_atom_cyl}

So far we have considered the retarded limit where the distance $d \gg d_{10}$ is much bigger than the retardation length $d_{10}=c/\omega_{10}$ set by the transition frequency of a two-state atom (see App.~\ref{Appendix} for its polarizability). Next we shall assume the opposite limit where $d\ll d_{10}$ which can be realized to leading order by taking the limit $c\to\infty$. The resulting interaction is known as London force. In the following we consider finite temperatures so that to leading order in $d/d_{10}$ one has
\begin{equation}
\label{E_London_finite_T}
 E^{L}_{T} = - \lim_{c\to\infty}\kT{\sum_{n=0}^{\infty}}'\Tr{{\mathcal{N}}(\kappa_{n})} \, .  
\end{equation} 
Here for need to substitute for the polarizability of the atom the frequency dependent expression of Eq.~\eqref{eq:polarizability_two_state}. Then the matrix $\mathcal{N}(\kappa_{n})$ depends on the combinations $d\kappa_n$, $R_c \kappa_n$ and $d_{10}\kappa_n= 2\pi n k_B T/(\hbar \omega_{10})$. The first two combinations scale as $1/c$ and hence tend to zero whereas the latter obviously remains finite for $c\to\infty$. Hence the non-retarded interaction is given by the Matsubara sum taken over two times the classical interaction energy $E_{cl}$ of Eq.~\eqref{E_Casimir_atomo_interno_a_cilindro_high_T} with $\alpha_P$ replaced by $\alpha_P/[1+(2\pi n k_B T/(\hbar \omega_{10}))^2]$. The sum over $n$ can be carried out easily, and one obtains the following result for the London interaction valid for all temperatures 
\begin{equation} 
\label{eq:London_limit_general}
E^{L}_{T}(\delta) = \frac{\hbar\omega_{10}}{2\kT}\coth\left(\frac{\hbar\omega_{10}}{2\kT}\right)E_{cl}(\delta).  
\end{equation} 
This shows that the effect of geometry (curvature) in the non-retarded limit is fully determined by the
geometry dependence in the classical limit.
Note that, in the zero and high temperature limits of the London energy we have 
\begin{equation} 
\label{eq:London_limit_0+cl}
E^{L}_{0}(\delta) = \frac{\hbar\omega_{10}}{2\kT}E_{cl}(\delta), \quad E^{L}_{cl}(\delta) = E_{cl}(\delta).  
\end{equation}

\subsection{Retarded Casimir-Polder energy of an atom outside a cylinder}

\subsubsection{Zero temperature limit}

The Casimir potential at zero temperature of an atom outside a perfectly conducting cylinder at a distance $d$ from the cylinder axis can be written as
\begin{equation}
\label{eq:E_atom_outside}
E_{0} = - \frac{\hc}{R_c^{4}}\left(g_{0}^{E}(\delta)\alpha_{E} + g_{0}^{M}(\delta)\alpha_{M}\right),
\end{equation}
where $\delta = d/R_c$. The dependence on $\delta$ is determined by the integrals
\begin{equation}
\label{AuxK}
\mathcal{K}_{\alpha,m}^{P}(\delta) = \int_{0}^{\infty}dt\,t^{\alpha}\sum_{n\in\mathbb{Z}}
 K_{n+m}^{2}(\delta t)\mathcal{T}_{C\,n,P;n,P}( t) \, ,
\end{equation}
where $\mathcal{T}_C$ is the diagonal T--matrix that described the scattering outside the cylinder.  The matrix elements are given by
\begin{align}
\mathcal{T}_{C \,n,M;n,M}(t) & = - \frac{I'_{n}(t)}{K'_{n}(t)}\, , \\
\mathcal{T}_{C\,n,N;n,N}(t) & = - \frac{I_{n}(t)}{K_{n}(t)}\, .
\end{align}
Then the amplitudes $g_{0}^{E}(\delta)$ and $g_{0}^{M}(\delta)$ are given by
\begin{align}
\label{gE0}
g_{0}^{E}(\delta) & = \frac{1}{4\pi}\left( \mathcal{K}_{3,1}^{M}(\delta) - \mathcal{K}_{3,1}^{N}(\delta) - 2 \mathcal{K}_{3,0}^{N}(\delta) \right) \, , \\
\label{gM0}
g_{0}^{M}(\delta) & = - \frac{1}{4\pi}\left( \mathcal{K}_{3,1}^{M}(\delta) - \mathcal{K}_{3,1}^{N}(\delta) + 2 \mathcal{K}_{3,0}^{M}(\delta) \right) \, .
\end{align}
Using again the uniform asymptotic expansion for Bessel functions, we find for {\it small}  atom-surface distance $\tilde\ell=d-R_c$ the expansion
\begin{align}
\label{short_distance_limit_gE0}
g_{0}^{E} & = \frac{3}{8\pi} \left(\frac{R_c}{\tilde\ell}\right)^4 - \frac{13}{60\pi} \left(\frac{R_c}{\tilde\ell}\right)^3 + \frac{311}{1680\pi}\left(\frac{R_c}{\tilde\ell}\right)^2 + \cdots \, ,\\
\label{short_distance_limit_gM0}
g_{0}^{M}  & = -\frac{3}{8\pi} \left(\frac{R_c}{\tilde\ell}\right)^4 + \frac{17}{60\pi} \left(\frac{R_c}{\tilde\ell}\right)^3 - \frac{431}{1680\pi}  \left(\frac{R_c}{\tilde\ell}\right)^2 + \cdots \, .
\end{align}
which is consistent with \cite{PFA_Atom_Cylinder}. Comparison with Eqs.~\eqref{short_distance_limit_fE0}, \eqref{short_distance_limit_fM0} shows that the interior and exterior cases are related by $R_c \to -R_c$. While an interior atom sees negative surface curvature, an exterior atom experiences positive curvature.

\begin{figure}[h]
\begin{center}
\includegraphics[scale=1]{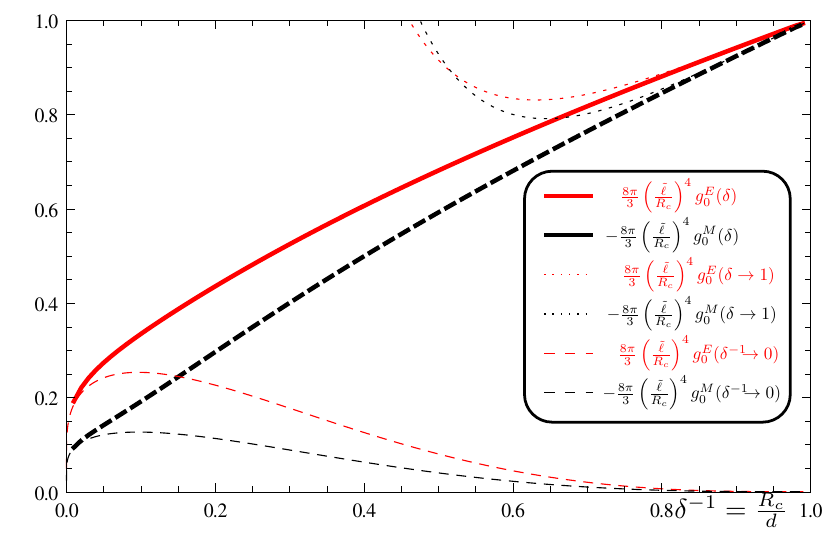}
\caption{(Color online). Amplitudes of the exterior Casimir-Polder interaction between an atom and a cylindrical shell as function of $1/\delta = R_{c}/d$. Shown are the limiting analytical results for short distances (dotted curves, see Eqs.~\eqref{short_distance_limit_gE0}, \eqref{short_distance_limit_gM0}), and when the atom is far away from the cylinder (dashed curves, see Eqs.~\eqref{large_distance_limit_gE0}, \eqref{large_distance_limit_gM0}). The thick curves correspond to the full numerical result for $g_{0}^{E}(\delta)$ (Eq.~\eqref{gE0}) and $g_{0}^{M}(\delta)$ (Eq.~\eqref{gM0}).}
\label{zeroT_Retarded_Casimir_energy_atom_outside_PM_Cylinder}
\end{center}
\end{figure}

In the large distance limit, $d\gg R_{c}$, we obtain the following asymptotic amplitudes
\begin{align}
\label{large_distance_limit_gE0}
g_{0}^{E} & =  \frac{1}{3\pi\ln(d/R_{c})} \left(\frac{R_c}{d}\right)^4 \, , \\
\label{large_distance_limit_gM0}
g_{0}^{M} & = -  \frac{1}{6\pi\ln(d/R_{c})} \left(\frac{R_c}{d}\right)^4 \, ,
\end{align}
yielding for the Casimir potential
\begin{equation}
E_{0} = -\frac{\hc(2 \alpha_{E} - \alpha_{M})}{6 \pi \, d^{4}\ln(d/R_c)} \, .
\end{equation}
Numerical and analytical results are compared in Fig.~\ref{zeroT_Retarded_Casimir_energy_atom_outside_PM_Cylinder}, where the convergence of the numerical data to the various limits is shown.

\subsubsection{High temperature limit}

We consider now high temperatures so that $d \gg \lambda_{T}$. 
Similar to the interior case, the Casimir potential can be written as
\begin{equation}
\label{E_Casimir_atomo_externo_a_cilindro_high_T}
E_{cl}(r_{c}) = - \frac{\kT}{R_c^{3}}\left(g_{cl}^{E}(\delta)\alpha_{E} + g_{cl}^{M}(\delta)\alpha_{M}\right).
\end{equation}
The functions $g_{cl}^{E}(\delta)$ and $g_{cl}^{M}(\delta)$ can be written in terms of the functions $\mathcal{K}_{\alpha,m}^{P}(\delta)$ as
\begin{align}
\label{gEcl}
g_{cl}^{E}(\delta) & = - \frac{1}{\pi}\left( \mathcal{K}_{2,1}^{N}(\delta) + \mathcal{K}_{2,0}^{N}(\delta) \right) \, , \\
\label{gMcl}
g_{cl}^{M}(\delta) & = - \frac{1}{\pi}\left( \mathcal{K}_{2,1}^{M}(\delta) + \mathcal{K}_{2,0}^{M}(\delta) \right) \, .
\end{align}
The asymptotic expansion of Bessel functions yields again the asymptotic behavior of $\mathcal{K}_{\alpha,m}^{P}$ when the atom is close to the surface of the cavity. 
In this limit we obtain
\begin{align}
\label{short_distance_limit_gEcl}
g_{cl}^{E} &= \frac{1}{4} \left(\frac{R_c}{\tilde\ell}\right)^3 - \frac{1}{8} \left(\frac{R_c}{\tilde\ell}\right)^2 + 
\frac{1}{8} \left(\frac{R_c}{\tilde\ell}\right) + \cdots \, ,\\
\label{short_distance_limit_gMcl}
g_{cl}^{M} &= - \frac{1}{4} \left(\frac{R_c}{\tilde\ell}\right)^3 + \frac{1}{4} \left(\frac{R_c}{\tilde\ell}\right)^2 
 - \frac{1}{4} \left(\frac{R_c}{\tilde\ell}\right) + \cdots \, .
\end{align}
This results is again related to the corresponding one for the interior case in Eqs.~\eqref{short_distance_limit_fEcl},~\eqref{short_distance_limit_fMcl} by $R_c \to -R_c$. 

\begin{figure}[h]
\begin{center}
\includegraphics[scale=1]{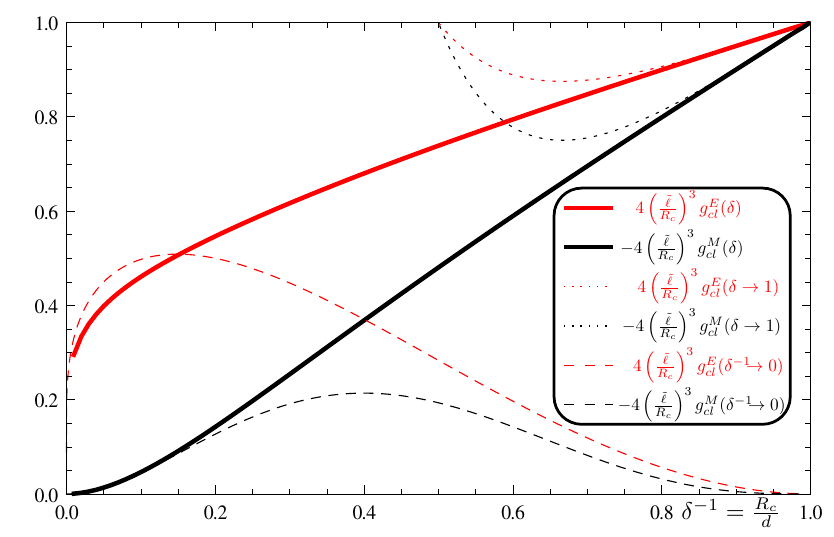}
\caption{(Color online). Equivalent of Fig.~\ref{zeroT_Retarded_Casimir_energy_atom_outside_PM_Cylinder}
for the classical limit. Shown are the analytical results for short surface distances $\tilde\ell$ (dotted curves, see Eqs.~\eqref{short_distance_limit_gEcl}, \eqref{short_distance_limit_gMcl}), and when the atom is far away from the cylinder (dashed curves, see Eqs.~\eqref{large_distance_limit_gEcl},~\eqref{large_distance_limit_gMcl}). The thick curves represent the full numerical result for 
$g_{cl}^{E}(\delta)$ (Eq.~\eqref{gEcl}) and $g_{cl}^{M}(\delta)$ (Eq.~\eqref{gMcl}).}
\label{highT_Retarded_Casimir_energy_atom_outside_PM_Cylinder}
\end{center}
\end{figure}

In the large distance limit, $d\gg R_{c}$, we find the asymptotic expressions
\begin{align}
\label{large_distance_limit_gEcl}
g_{cl}^{E} & =  \frac{\pi}{8\ln(d/R_{c})} \left(\frac{R_c}{d}\right)^3 \, , \\
\label{large_distance_limit_gMcl}
g_{cl}^{M} & = - \frac{63\pi}{128}\left(\frac{R_c}{d}\right)^5 \, ,
\end{align}
which yield the Casimir potential
\begin{equation}
E_{cl} = -\frac{\kT\pi\alpha_{E}}{8 d^{3}\ln(d/R_{c})}.
\end{equation}
Note that the leading asymptotic interaction is dominated by the electric response of the atom only.
The results of all full numerical computation of the interaction at all distances is shown in 
Fig.~\ref{highT_Retarded_Casimir_energy_atom_outside_PM_Cylinder} together with the analytically studied limits.

\subsection{Non-retarded London energy of an atom outside the cylinder}

In the non-retarded limit with $d \ll d_{10}$ for the exterior case of an atom outside the cylinder exactly the same relations as in the interior case hold, see Eqs.~\eqref{eq:London_limit_general}, \eqref{eq:London_limit_0+cl}.

\section{Casimir energy of a metal sphere inside a metal Cylinder}

\subsection{Proximity force approximation}
\label{PFA of the Casimir energy of a sphere inside a metal Cylinder}

In this section we come back to an interior situation. Instead of an atom, we place a macroscopic metallic sphere inside the cylinder. We assume that both the cylinder and the sphere are perfectly conducting, and have radii $R_c$ and $R_s$, respectively. The sphere--center to cylinder--axis separation is $d\le R_c-R_s$. Before computing the exact interaction in the next section, we consider here the proximity force approximation (PFA). In general, the PFA energy $E_{\PFA}$ for two surfaces is given by
\begin{equation}\label{eq:pfa-general}
E_{\PFA} = \int dA\, E_{\parallel}(h)\,,
\end{equation}
where $E_{\parallel}(h)$ is the energy per unit area for two parallel plates 
of distance $h$. In the above expression the integration is performed along one surface with $h$ the local distance to the other surface. After integration, the result is expanded for small distances, and only the leading order is retained. To this order, the precise direction along which $h$ is measured, is unimportant. 

Assuming the origin at the axis of the cylinder (oriented along the $x$-axis), the position of a surface element on the cylinder is $(x_{c},  y_{c}=R_{c}\sin(\phi_{c}), z_{c}=R_{c}\cos(\phi_{c}))$, and the position of a surface element on the sphere is 
$(x_{s}=R_{s}\sin(\theta_{s})\cos(\phi_{s}), y_{s}=R_{s}\sin(\theta_{s})\sin(\phi_{s}), z_{s}=d+R_{s}\cos(\theta_{s}))$. The distance between the two surface elements then is
\begin{equation}
\label{eq:distance-h}
h= z_{c} -z_{s} \,.
\end{equation}
The center-to-axis distance $d$ is related to the minimal surface-to-surface distance $\ell$ between the sphere and the cylinder by $d = R_{c} -(\ell + R_{s})$. Hence Eq.~\eqref{eq:distance-h} can be written as
\begin{equation}\label{eq:distance-h-2}
h = \ell + R_{s}\left[ 1 - \cos(\theta_{s}) \right] - R_{c}\left[ 1 - \cos(\phi_{c}) \right]\,.
\end{equation}
Next we express $\phi_{c}$ in terms of $\theta_{s}$ and $\phi_{s}$, using $y_{c}=y_{s}$,
\begin{equation}\label{eq:angles-relation}
\sin(\phi_{c}) = \frac{R_{s}}{R_{c}}\sin(\theta_{s})\sin(\phi_{s})\,.
\end{equation}
Using Eq.~\eqref{eq:angles-relation} in Eq.~\eqref{eq:distance-h-2}
and making use of the fact that at short separations, the surface elements of the sphere and cylinder for which $\theta_{s}\ll 1$ and $\phi_{c}\ll 1$ contribute most to the interaction, the local distance $h$ can be approximated by
\begin{equation}\label{eq:distance-h-3}
h(\theta_{s},\phi_{s}) \approx \ell + \frac{R_{s}\theta_{s}^{2}}{2} 
\left[ 1 - \frac{R_{s}}{R_{c}}\sin^{2}(\phi_{s})\right]\,.
\end{equation}
We express Eq.~\eqref{eq:pfa-general} in terms of surface coordinates of the sphere,
\begin{equation}\label{eq:pfa}
{E}_{\PFA} = R_{s}^2\int_{0}^{2\pi}d\phi_{s}\int_{0}^{\pi}d\theta_{s}\sin \theta_{s}\, E_{\parallel}(h)\,,
\end{equation}
For small $\theta_s$ we use $\sin \theta_s\approx \theta_s$ and change the 
integration variable $\theta_s$ to $H$ which is defined by the right hand side of 
Eq.~\eqref{eq:distance-h-3} so that
\begin{equation}
\theta_{s}\,d\theta_{s}=\frac{dH} {R_{s}\big[1-\frac{R_{s}}{R_{c}} \sin^2(\phi_{s})\big]}\, .
\end{equation}
This yields
\begin{equation}
{E}_\mathrm{PFA}= R_{s}
\left[
\int _0^{2\pi}\frac{d\phi_{s}}{1-\frac{R_{s}}{R_{c}}\sin^2(\phi_{s})}
\right]
\int_{\ell}^{\infty}{d}H\, E_{\parallel}(H)\,,
\end{equation}
where we have moved the upper integration limit for $H$ to infinity since this does not change the leading behavior of the integral for small $\ell$. 
The integration over $\phi_s$ can be carried out easily and we obtain the PFA energy 
\begin{equation}\label{PFA_forallT}
{E}_\mathrm{PFA}= \frac{2\pi R_{s}}{\sqrt{1-\frac{R_{s}}{R_{c}}}}
\int_{\ell}^{\infty}{d}H\, E_{\parallel}(H)\,.
\end{equation}
In particular, for perfect conductors, we obtain 
\begin{equation}\label{PFA_quantum}
{E}_\mathrm{PFA,0} = - \hc\frac{R_{s}}{\ell^{2}}\frac{\pi^{3}}{720\sqrt{ 1 - \frac{R_{s}}{R_{c}}}}
\end{equation}
for zero temperature, and 
\begin{equation}\label{PFA_thermal}
{E}_\mathrm{PFA,cl} = - \kT\frac{R_{s}}{\ell}\frac{\zeta(3)}{4\sqrt{ 1 - \frac{R_{s}}{R_{c}}}}
\end{equation}
in the classical high temperature limit with $d\gg\lambda_T$.

\subsection{Numerical result}
\label{Casimir energy of a sphere inside a metal Cylinder - Numerical study}

\begin{figure}
\begin{center}
\includegraphics[width=1.0\columnwidth]{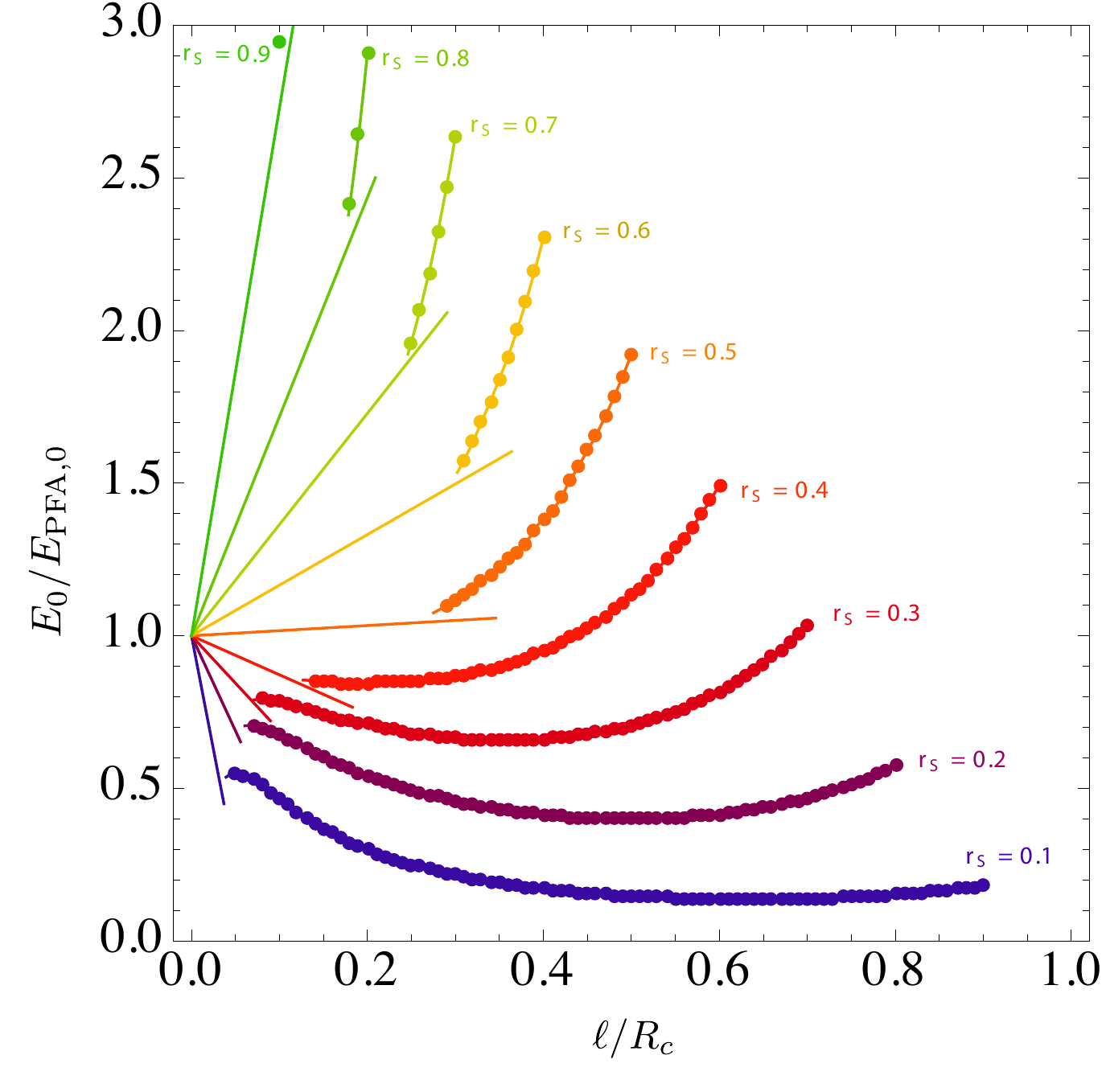}
\caption{(Color online) Numerical result for the zero temperature Casimir energy (normalized to PFA) for a sphere inside a cylinder for different ratios of the radii, ranging from $r_{s} = R_{s}/R_{c} = 0.1$ (bottom blue) to $r_{s} = 0.9$ (top green) in steps of $0.1$, as a function of $\ell/R_{c} < 1-r_s$. The 
curves terminate at the positions where the sphere is located at the axis of the cylinder ($\ell/R_{c} = 1-r_s$). The straight lines originating from unity represent the first correction to PFA, see Eq.~\eqref{eq:PFA_correction}.}
\label{Ratio_PFA_numerics_zeroT}
\end{center}
\end{figure}

\begin{figure}
\begin{center}
\includegraphics[width=1.\columnwidth]{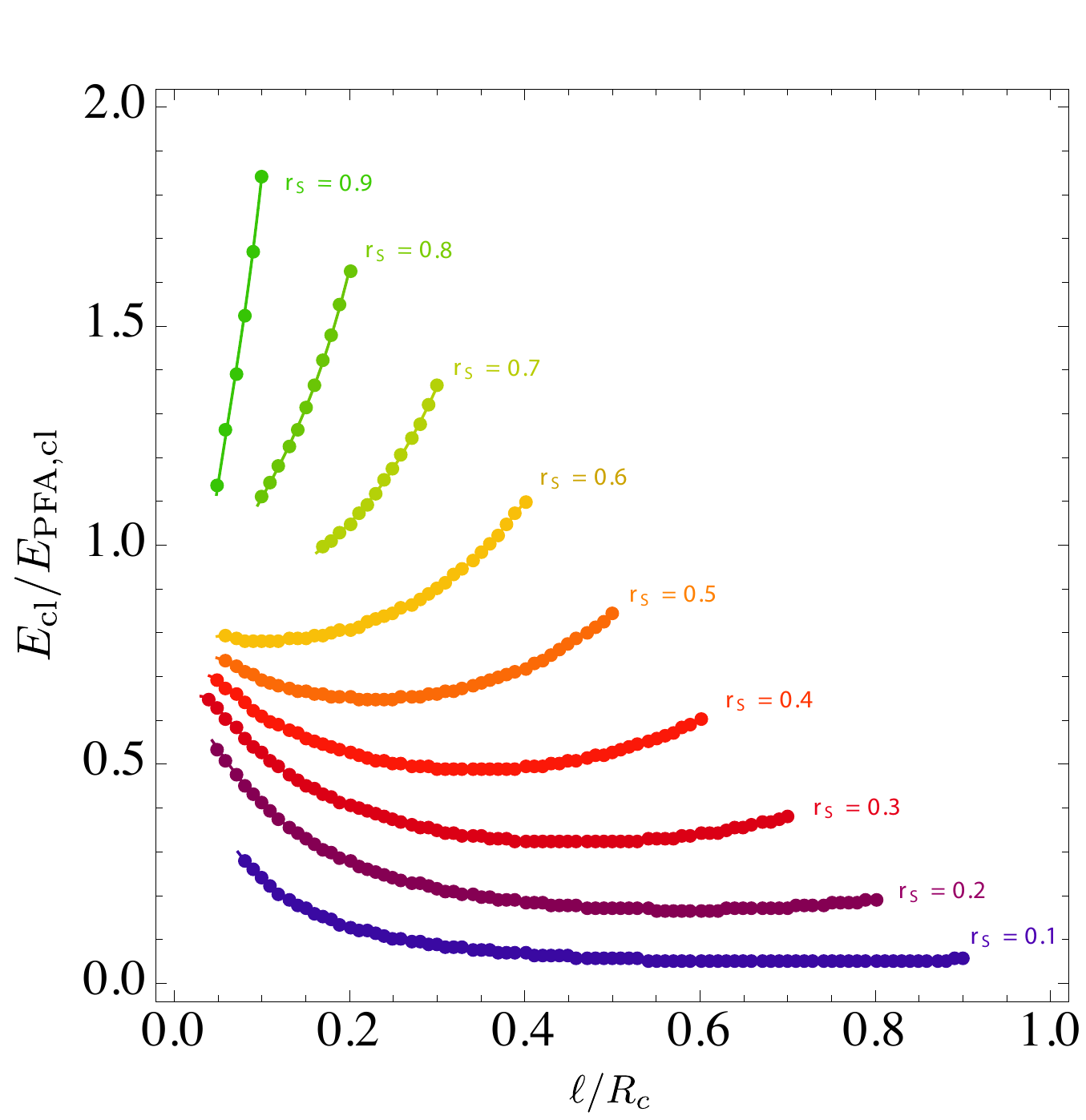}
\caption{(Color online) Same as Fig.~\ref{Ratio_PFA_numerics_zeroT} for the classical high temperature limit.}
\label{Ratio_PFA_numerics_highT}
\end{center}
\end{figure}

Now we compute the interaction for the geometry of the previous section numerically at all distances, using the scattering approach. Again, we consider zero temperature and the classical high temperature limit.  We obtain the Casimir energy numerically from the matrix $\mathcal{N}$ defined in Eq.~\eqref{Nmatrix_sphere_in_cylindrical_cavity} using Eqs.~\eqref{zero T Casimir energy}, \eqref{high T Casimir energy}. The T-matrix elements for a perfectly conducting cylinder and sphere are given in App.~\ref{Appendix}.  In order to perform the numerical calculations, we have restricted the vector spherical multipoles to $l \leq 20$ in the zero temperature case and to $l \leq 30$ in the high temperature case. The cylindrical multipoles are limited to $|n| \leq 500$.  The results are shown in Fig.~\ref{Ratio_PFA_numerics_zeroT} ($T=0$) and Fig.~\ref{Ratio_PFA_numerics_highT} (classical limit)  as function of $\ell/R_c < 1-R_s/R_c$ for different radii of the sphere.  The data for the numerically computed Casimir 
energy are normalized to the PFA [see Eqs.~\eqref{PFA_quantum}, \eqref{PFA_thermal}].

The multipole expansion for Casimir interactions works best at large separations, and an increasing number of multipoles is required when the surfaces of the bodies approach each other. This is clearly visible in the numerical data that should converge to unity for $\ell \to 0$ when normalized to the PFA but fail to do so below $\ell/R_c \lesssim 0.05$. However, for $\ell/R_c \gtrsim 0.05$ we expect our numerical results to be reliable, and they clearly show strong deviations from the PFA when the sphere is moved towards the axis of the cylinder (reached for $\ell/R_c=1-R_s/R_c$). The sign of the corrections to the PFA depend on the ratio of the radii, $r_s=R_s/R_c$, in both the zero and the high temperature limit.

Corrections to the PFA can be computed in some cases by a derivative expansion \cite{Derivative_Expansion_PFA_energy}. This expansion can be used in the limit where the radii of curvature of the surfaces are much larger than their shortest distance, i.e., in the present geometry for $R_s$, $R_c \gg \ell$. By using the expansion for perfect conductors at $T=0$, we obtain for the Casimir interaction at small $\ell \ll R_c$,
\begin{widetext}
\begin{equation}
  \label{eq:PFA_correction}
  \mathcal{E}_{0} = \mathcal{E}_\mathrm{PFA,0} 
\left[ 1 + \left( -1 -\frac{3}{2} r_s + r_s\sqrt{1-r_s} +\frac{2\beta+(2-3\beta)r_s+\beta r_s^2}
{1-r_s} \right) \frac{1}{r_s} \frac{\ell}{R_c} + \ldots\right] 
\end{equation}
\end{widetext}
with $\beta=(2/3)(1-15/\pi^2)$\cite{Derivative_Expansion_PFA_energy}.  The first correction to the PFA vanishes for $r_s=0.4881$ where it changes sign which is consistent with the numerical results in Fig.~\ref{Ratio_PFA_numerics_zeroT}. The numerical results approach the prediction of the derivative expansion at small surface-to-surface distance $\ell$ better for small $r_s$. Presumably, for larger $r_s$ more multipoles need to be included in the numerical evaluation to obtain sufficient accuracy at short distances.
In the classical limit, there exists no derivative expansion for perfect metals due to a non-analytic behavior of the kernel for small momenta \cite{Fosco_2012a}.

\section{Casimir interaction of two atoms inside a cylindrical cavity}
\label{Effect of confinement and non--pairwise behavior of Casimir energy}

It is widely known that Casimir interactions are not pairwise additive: The energy of a system of a given number of objects is not the sum of the Casimir energies of all pairs.  As a consequence, new phenomena can appear, as the non--monotonicity of the Casimir force for two cylinders~\cite{SJohnson}, and for spheres or atoms in the presence of a perfect metal plate~\cite{Emig_Pablo}. Similar effects can be expected for the interaction of objects inside cavities due to the confinement of field fluctuations.
In particular, it is known that the Casimir force between two atoms which are confined between two dielectric parallel plates increases several orders of magnitude~\cite{2_atoms_in_dielectric_capacitor}. When the dielectric plates are replaced by perfect metal plates, the effect is even more pronounced: At zero temperature, the Casimir potential of the atoms decays no longer  $\sim d^{-7}$ as in free space but  $\sim d^{-5}$~\cite{2_atoms_in_PM_capacitor_zeroT}. In the high temperature limit, the Casimir potential was found to decay exponentially~\cite{2_atoms_in_PM_capacitor_highT}. It is expected that the confinement effects increase with the degree of spatial confinement so that a cavity, e.g., a cylindrical shell should produce more pronounced consequences. 

In this section we study the Casimir potential for two atoms inside a perfectly conducting cylindrical shell. We assume that the atoms have both electric and magnetic dipole polarizabilities. The Casimir energy of
this 3-body problem is given by Eq.~\eqref{Energy_T_finite} with the block matrix
\begin{equation}
\mathcal{I} - \mathcal{N} = \left(\begin{array}{ccc}
   \mathcal{I} & - \mathcal{T}_{A_1}\mathcal{X}_{A_1A_2} & - \mathcal{T}_{A_1}\mathcal{X}_{A_1C}\\
 - \mathcal{T}_{A_2}\mathcal{X}_{A_2A_1} &   \mathcal{I} & - \mathcal{T}_{A_2}\mathcal{X}_{A_2C}\\
 - \mathcal{T}_{C}\mathcal{X}_{CA_1} & - \mathcal{T}_{C}\mathcal{X}_{CA_2} & \mathcal{I}
\end{array}\right) \, ,
\end{equation}
where the matrix labels $A_1$, $A_2$ and $C$ stand for atom $1$, atom $2$, and the cylindrical shell.
For example, the matrix $\mathcal{X}_{CA_1}$ describes the translation between the cylindrical shell and atom $1$. The determinant of this matrix can be rearranged using the relations
\begin{eqnarray}
\Det{\begin{array}{cc}
A & B\\
C & D
\end{array}} & = & \Det{A}\Det{ D - C\,A^{-1}\,B }\nonumber\\
& = & \Det{D}\Det{ A - B\,D^{-1}\,C } \, .
\end{eqnarray}
The Casimir energy at zero temperature can then be written as
\begin{widetext}
\begin{equation}
\label{eq:E3}
E_{3} = \frac{\hc}{2\pi}\int_{0}^{\infty}d\kappa \log \left\{ \Det{\mathcal{I} - \mathcal{N}_{A_1C}}\Det{\mathcal{I} - \mathcal{N}_{A_2C}}\Det{\mathcal{I} - \mathcal{R}_{A_2A_1}} \right\}
\end{equation}
where
\begin{align}
\mathcal{N}_{A_1C} & = \mathcal{T}_{A_1}\mathcal{X}_{A_1C}\mathcal{T}_{C}\mathcal{X}_{CA_1}\, ,\\
\mathcal{N}_{A_2C} & = \mathcal{T}_{A_2}\mathcal{X}_{A_2C}\mathcal{T}_{C}\mathcal{X}_{CA_2}\, ,\\
\label{R_Matrix_2_objects_inside_a_cylindrical_cavity}
\mathcal{R}_{A_2A_1} & = \mathcal{T}_{A_1}\left( \mathcal{X}_{A_1A_2} + \mathcal{X}_{A_1C}\mathcal{T}_{C}\mathcal{X}_{CA_2} \right)\left(\mathcal{I} - \mathcal{N}_{A_2C}\right)^{-1}
 \mathcal{T}_{A_2}\left( \mathcal{X}_{A_2A_1} + \mathcal{X}_{A_2C}\mathcal{T}_{C}\mathcal{X}_{CA_1} \right)\left( \mathcal{I} - \mathcal{N}_{A_1C} \right)^{-1}\, .
\end{align}
The first two matrices describe the interaction of each atom with the cavity, and the third matrix describes the interaction between the two atoms, taking into account the presence of the cavity.  In the absence of the cavity, the latter matrix reduces to $\mathcal{R}_{A_2A_1} = \mathcal{T}_{A_1}\mathcal{X}_{A_1A_2}\mathcal{T}_{A_2}\mathcal{X}_{A_2A_1}$ which describes  two atoms in free space. 

It is convenient to express the matrix $\mathcal{R}_{A_2A_1}\equiv \mathcal{R}$ in spherical multipole basis which reads
\begin{equation}
\mathcal{R}_{l m,l'' m''} = \mathcal{T}_{A_1\,l m,l m}\tilde{\mathcal{U}}_{A_1A_2\, l m,l' m'}(d,\textbf{X}_{12})\mathcal{T}_{A_2\,l' m',l' m'}\tilde{\mathcal{U}}_{A_2 A_1\,l' m',l'' m''}(d,\textbf{X}_{21}),
\end{equation}
where here and in the following repeated indices are summed over. Here $d$
is the distance between the atoms and the axis of the cylindrical cavity, and $\tilde{\mathcal{U}}_{A_\alpha A_\beta\,l m,l' m'}(d,\textbf{X}_{\alpha\beta})$ are the modified translation matrices for the translation by the vector $\textbf{X}_{\alpha\beta}\equiv \pm h \hat{\bf z}$  along the cylinder axis from atom $\alpha$ to atom $\beta$. They are defined by
\begin{align}
\tilde{\mathcal{U}}_{A_\alpha A_\beta\,l m,l' m'}(d,\textbf{X}_{\alpha\beta}) &=\left( \mathcal{U}_{A_\alpha A_\beta\, l m,l'' m''}(\textbf{X}_{\alpha\beta}) + D_{l m,n k_{z}}\mathcal{V}_{A_\alpha C \,n k_{z}, n' k_{z}}\mathcal{T}_{C\,n' k_{z},n' k_{z}}\mathcal{V}^{\dagger}_{C A_\beta \,n' k_{z}, n'' k_{z}}D^\dagger_{n'' k_{z},l'' m''}\frac{C^{c}}{C^{s}} \right) \nonumber \\
&\times \left(\mathcal{I} -\mathcal{N}_{A_\beta C}\right)^{-1}_{l'' m'',l' m'} \, ,
\end{align}
\end{widetext}
where $\mathcal{N}_{A_\beta C}=\mathcal{T}_{A_\beta}\mathcal{X}_{A_\beta C}\mathcal{T}_C\mathcal{X}_{CA_\beta}$ describes the Casimir interaction of an atom with the cylindrical cavity, and the constants $C^c$ and $C^s$ are defined below Eq.~\eqref{Sph_2_Cyl}. 
The translation matrices $\mathcal{U}$ are defined in App.~\ref{Appendix}.
Note that the polarization indices and the integral over $k_z$ are not shown explicitly. 

In the following, we are interested in the interaction between the two atoms (and not the change of energy when the atoms are moved away from the cylinder axis). This interaction is given by the last determinant of Eq.~\eqref{eq:E3} and hence by the matrix $\mathcal{R}$.
Under the two assumptions formulated in the beginning of Sec.~\ref{Casimir energy of an atom inside a metallic Cylinder}, we need to retain only the contribution linear in $\mathcal{R}$ so that the Casimir energy is given by
\begin{equation}
\label{Energy_2_atoms_in_cavity}
E_{T} = \kT{\sum_{n=0}^{\infty}}'\Tr{\tilde{\mathcal{R}}(\kappa_{n})}\, ,
\end{equation}
where $\tilde{\mathcal{R}}$ is defined by the (symbolic) expression
\begin{align}
\label{R_Matrix_2_atoms_inside_a_cylindrical_cavity}
\tilde{\mathcal{R}} & = 
\mathcal{T}_{A_2}\left( \mathcal{U}_{A_2A_1} + D\mathcal{V}_{A_2C}\mathcal{T}_{C}\mathcal{V}^{\dagger}_{CA_1}D^{\dagger} \right)\nonumber\\
&  \times\mathcal{T}_{A_1}\left( \mathcal{U}_{A_1A_2} + D\mathcal{V}_{A_1C}\mathcal{T}_{C}\mathcal{V}^{\dagger}_{CA_2}D^{\dagger} \right),
\end{align}
that is the simplification of $\mathcal{R}$ correct to linear order in the polarizabilities of the atoms.  The matrix $\tilde{\mathcal{R}}$ has four contributions that can be understood as distinct scattering processes. The two terms containing $\mathcal{U}$ describe direct scattering between the two atoms not involving the cylindrical cavity. The other two terms containing $\mathcal{T}_{C}$ correspond to indirect scattering between the atoms with a reflection at the cylindrical cavity. 

In the following, we consider the zero and high temperature classical limit of the interaction between the atoms that are located on the cylinder axis and have separation $h$. In these two limits, the Casimir interaction energies can be written as
\begin{align}
\label{ZeroT_energy_2a_1PMC}
E_{0} & = - \frac{\hc}{4\pi R_{c}^{7}} \left[ e^{EE}_{0}(\eta)\alpha^{E}_{1}\alpha^{E}_{2} + e^{MM}_{0}(\eta)\alpha^{M}_{1}\alpha^{M}_{2} + \right.\nonumber\\
 & \left. + e^{EM}_{0}(\eta)\left(\alpha^{E}_{1}\alpha^{M}_{2} + \alpha^{M}_{1}\alpha^{E}_{2}\right) \right],\\
\label{HighT_energy_2a_1PMC}
E_{cl} &= - \frac{\kT}{R_{c}^{6}} \left[ e^{EE}_{cl}(\eta)\alpha^{E}_{1}\alpha^{E}_{2} + e^{MM}_{cl}(\eta)\alpha^{M}_{1}\alpha^{M}_{2}\right] \, ,
\end{align}
respectively, with $\eta = h/R_c$. Here $\alpha_j^P$ are the dipolar electric ($P=E$) and magnetic ($P=M$) polarizabilities of the atoms. Note that there are no terms proportional to $\alpha^{E}_{i}\alpha^{M}_{j}$ in the high temperature limit. Below, we shall give analytical and numerical results for the amplitude functions $e_0$ and $e_{cl}$ in various limiting cases.

\subsection{Retarded limit}

We first consider the situation where the distance between the atoms is much bigger than the retardation length, $h \gg d_{10}$. This corresponds to the so-called Casimir--Polder or retarded limit. 

\subsubsection{Zero temperature}

Indeed, when thermal effects are unimportant  and the distance $h \ll R_c$, the atoms interact as in free space, and we obtain the usual Casimir--Polder potential for two atoms  at zero temperature, corresponding to
\begin{equation}\label{eEE_eMM_0_free}
e^{EE}_{0}(\eta) = e^{MM}_{0}(\eta) = \frac{23}{\eta^{7}},
\end{equation}
\begin{equation}\label{eEM_0_free}
e^{EM}_{0}(\eta) = - \frac{7}{\eta^{7}} \,.
\end{equation}
In the opposite limit where $h \gg R_c \gg d_{10}$ the atoms experience the confinement by the cylindrical shell,
leading to an exponential decay of their interaction. For $\eta\gg 1$, we obtain the limiting amplitude functions
\begin{align}
\label{eMM_0_as}
e^{MM}_{0} & = \frac{\sqrt{\pi^{9}{j'_{1,1}}^{21}}{Y_{1}'}^{4}(j'_{1,1})}{8\left( {j'_{1,1}}^{2} - 1 \right)^{2}}\frac{e^{-2j'_{1,1}\eta}}{\eta^{1/2}} = 286.20 \frac{e^{-2j'_{1,1}\eta}}{\eta^{1/2}}\, ,\\
\label{eEE_0_as}
e^{EE}_{0} & = \frac{3\sqrt{\pi^{9}{j'_{1,1}}^{17}}{Y_{1}'}^{4}(j'_{1,1})}{32 \left( {j'_{1,1}}^{2} - 1 \right)^{2}}\frac{e^{-2j'_{1,1}\eta}}{\eta^{5/2}}  = 63.32 \frac{e^{-2j'_{1,1}\eta}}{\eta^{5/2}} \, , \\
\label{eEM_0_as}
e^{EM}_{0} & = -\frac{8\sqrt{2} \pi^{2} {j'_{1,1}}^{4} {Y_{1}'}^{2}(j'_{1,1})}{  {j'_{1,1}}^{2} - 1}\frac{e^{-j'_{1,1}\eta}}{\eta^{5}} =  - 189.59 \frac{e^{-j'_{1,1}\eta}}{\eta^{5}}\, ,
\end{align}
where $j'_{1,1}=1.84118$ is the lowest zero of the derivative of the Bessel functions of first kind, $J'_1(x)$, and $Y'_1(x)$ is the derivative of a Bessel function of second type. 
The electromagnetic fluctuations inside the confining cylinder are effectively massive with a mass $\sim R_c$ that produces the exponential decay of the interaction for $h \gg R_c$. 
\begin{figure}[th] 
\begin{center} 
\includegraphics{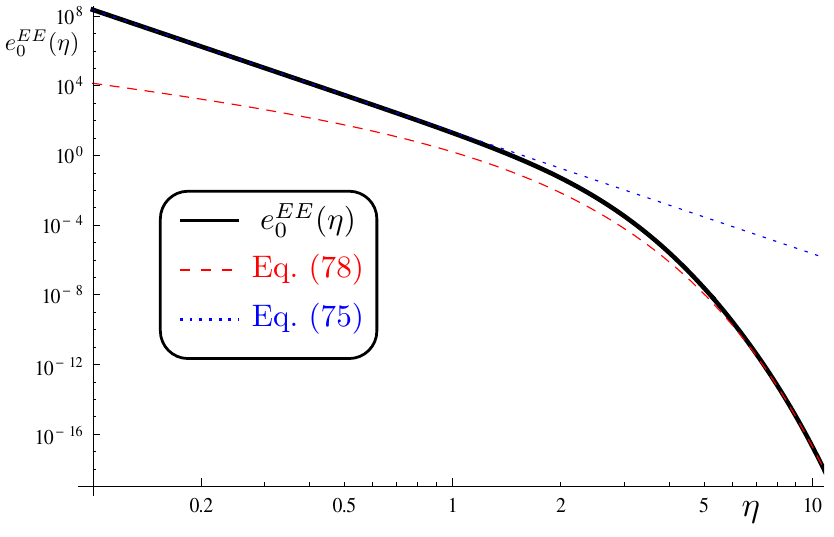} \includegraphics{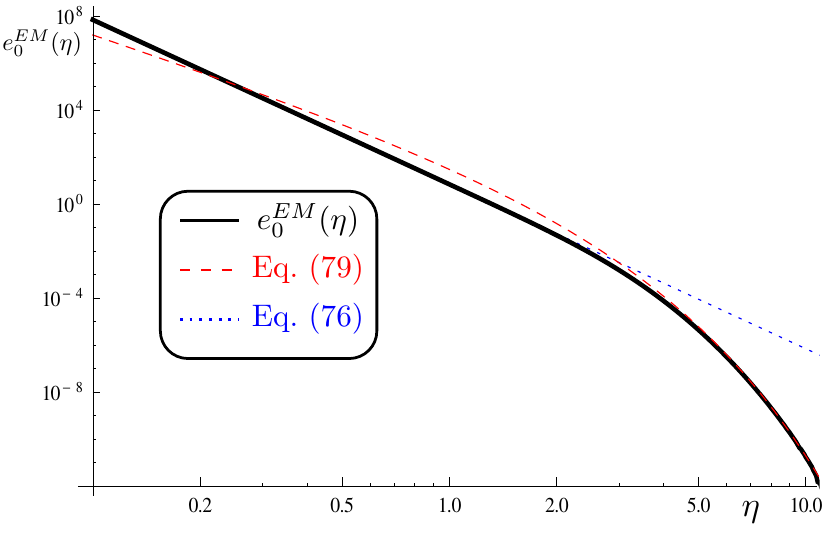} \includegraphics{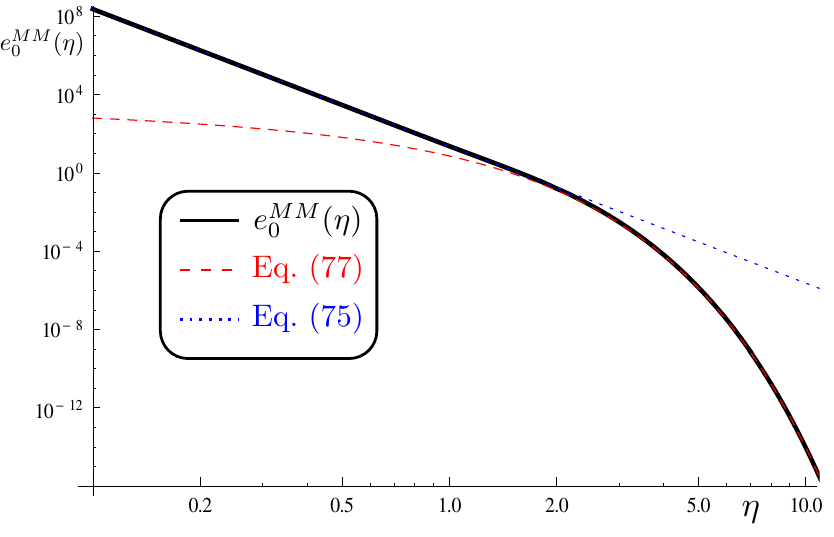} 
\caption{(Color online) Log--log plot of the numerical results for the functions $e_{0}(\eta)$ as a function of $\eta = h/R_{c}$ for the three combinations of polarizations (solid curves). The dashed curves represent the large distance limits ($\eta \gg 1$) given by Eqs.~\eqref{eMM_0_as}, \eqref{eEE_0_as} and \eqref{eEM_0_as}, and the dotted curves correspond to the short distance limits ($\eta \ll 1$) given by Eqs.~\eqref{eEE_eMM_0_free} and \eqref{eEM_0_free}).}
\label{ZeroTCasimir_energy_2_atoms_inside_PMCylinder_separated}
\end{center}
\end{figure}
We have evaluated the amplitude functions $e_{0}(\eta)$ for intermediate values of $\eta$ numerically. The results are shown in Fig.~\ref{ZeroTCasimir_energy_2_atoms_inside_PMCylinder_separated} for the three possible combinations of the polarizations. Shown in the plots are also the limiting analytical results of Eqs.~\eqref{eEE_eMM_0_free}, \eqref{eEM_0_free}, \eqref{eMM_0_as}, \eqref{eEE_0_as},  \eqref{eEM_0_as}. One can clearly observe the crossover between the two limiting expressions for $h \ll R_c$  and $h \gg R_c$.

\subsubsection{High temperature limit}
\label{HighT_Casimir_energy_2atoms_in_cavity}

Next, we consider the retarded regime in the high temperature classical limit. 
In this case, the electric and magnetic polarizations do not couple, as indicated already in Eq.~\eqref{HighT_energy_2a_1PMC}. For short distances, $h \ll R_c$, we recover the classical limit of the Casimir--Polder potential, corresponding to
\begin{equation}
\label{eEE_eMM_cl_free}
e^{EE}_{cl}(\eta) = e^{MM}_{cl}(\eta) = \frac{3}{\eta^{6}} \, .
\end{equation}
In the large distance limit, $h \gg R_{c} \gg d_{10}$, the potential decays again exponentially due to the confinement,
\begin{align}
\label{eEE_cl_as}
e^{EE}_{cl} & = \frac{\pi^{4}j_{0,1}^{6}Y_{0}^{4}(j_{0,1})}{8}e^{-2j_{0,1}\eta} = 159.23 e^{-2j_{0,1}\eta}\, , \\
\label{eMM_cl_as}
e^{MM}_{cl} & = \frac{\pi^{4}{j'_{1,1}}^{10} {Y'_{1}}^{4}(j'_{1,1})}{16\left({j'_{1,1}}^{2} - 1\right)^{2}}e^{-2j'_{1,1}\eta} = 59.50 e^{-2j'_{1,1}\eta}\, ,
\end{align}
where $j_{0, 1}=2.40483$ is the lowest zero of the Bessel function of the first kind $J_{0}(x)$.

\begin{figure}[h]
\begin{center}
\includegraphics{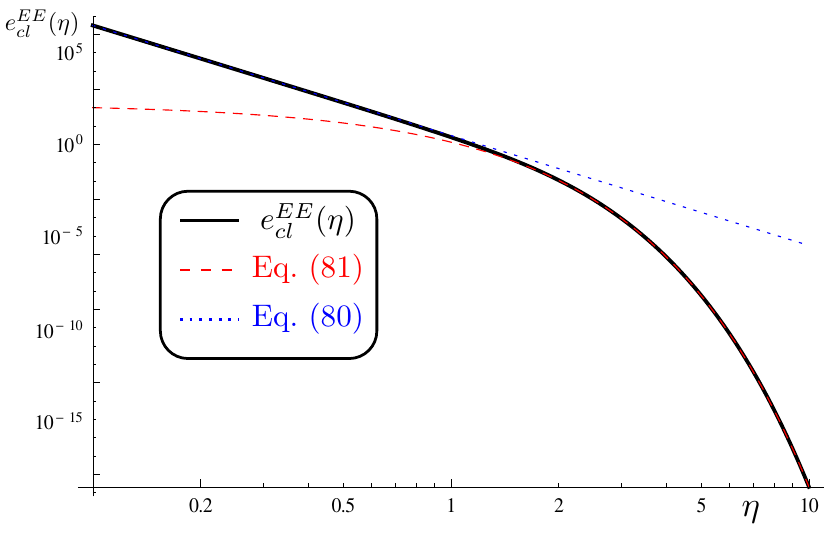}
\includegraphics{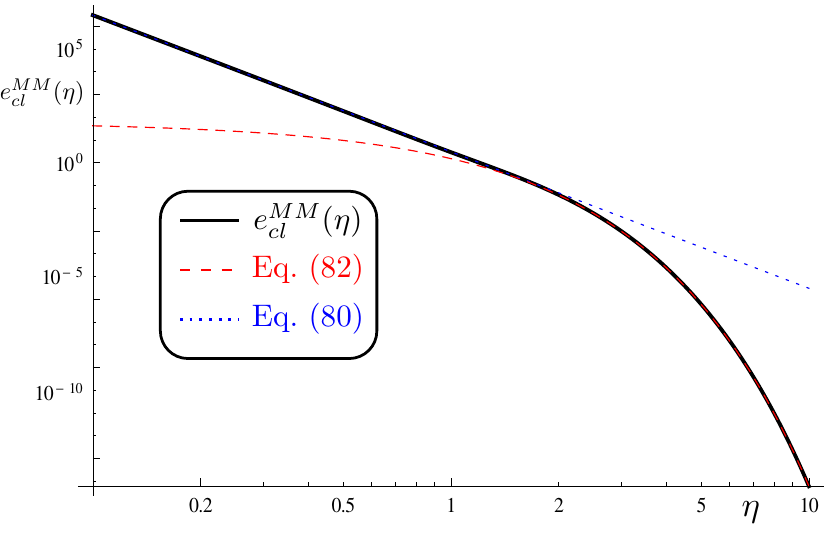}
\caption{(Color online) Log--log plot of the numerical results for the functions $e_{cl}(\eta)$ as a function of $\eta = h/R_{c}$ for the two combinations of polarizations (solid curves). The dashed curves represent the large distance limits ($\eta \gg 1$) given by Eqs.~\eqref{eEE_cl_as} and \eqref{eMM_cl_as}, and the dotted curves correspond to the short distance limits ($\eta \ll 1$) given by Eq.~\eqref{eEE_eMM_cl_free}}.
\label{HighTCasimir_energy_2_atoms_inside_PMCylinder_separated}
\end{center}
\end{figure}

To obtain information about intermediate distances, we have evaluated  the amplitude functions $e_{cl}(\eta)$ also numerically. The results are shown in Fig.~\ref{HighTCasimir_energy_2_atoms_inside_PMCylinder_separated}, together with the analytical results for large and small distances. The plots show that the limiting analytical results merge at $\eta$ of order unity and yield an accurate description of the interaction between the atoms for the entire range of separations.

\subsection{Non--retarded limit}

Finally, we consider the non--retarded limit where $h$, $R_c \ll d_{10}$ but $\eta=h/R_c$ is arbitrary. In this limit, the interaction between the atoms is usually referred to as London force. Formally, this case corresponds to an infinite velocity of light, $c\to\,\infty$. Here we consider arbitrary temperatures $T$.
Following the steps carried out in Sec.~\ref{sec:London_atom_cyl}, we can carry out the Matsubara sum and obtain the
London potential
\begin{equation}
\label{London_energy_2a_1PMC}
E^{L}_{T}(\eta) = f\left(\frac{\hbar\omega_{10}}{\kT}\right)E_{cl}(\eta),
\end{equation}
where the coefficient function $f(t)$ is given by
\begin{equation}
f(t) = \frac{t}{4}\frac{t + \sinh(t)}{\cosh(t)-1}
\end{equation}
and $E_{cl}(\eta)$ is the high temperature limit of the Casimir energy given by Eqs.~\eqref{HighT_energy_2a_1PMC}, \eqref{eEE_eMM_cl_free}.  Similar to the interaction between a single atom and a cylinder, the non-retarded interaction at finite temperature is related to the classical limit of the retarded interaction.
Note that in the zero and high temperature limit of the London potential we have
\begin{align}
E^{L}_{0}(\eta) & = \frac{\hbar\omega_{10}}{4\kT}E_{cl}(\eta) \, ,\\
E^{L}_{cl}(\eta) & = E_{cl}(\eta) \, .
\end{align}

\section{Discussion}
\label{Discussion}

We have studied how curvature and in particular confinement can effectively modify Casimir interactions between an atom and a curved surface and between atoms confined in perfectly conducting cylindrical cavities. For the interaction between a single atom and a cylindrical shell we have developed both large and short distance expansions that apply in the zero and high temperature limits, and for an interior and exterior atom. The expansions agree nicely with a full numerical evaluation at arbitrary atom-surfaces distances. Our results are relevant to understand the scattering of atoms at curved surfaces as, e.g., nanotubes. We have considered both the retarded and non-retarded limits. For the latter, we have shown that it is simply related to the classical (high temperature) limit of the Casimir-Polder potential. 

A problem with a higher degree of complexity is the interaction of a macroscopic spherical particle with a confining cylindrical shell. We have considered the limit of perfect conductivity for both the particle and the shell.  By computing the interaction numerically from a large number of mulipole moments, we were able to
compare to the proximity force approximation (PFA) and its correction as predicted by a gradient expansion.
We found nice agreement, confirming that the gradient expansion makes reliable predictions also for interior 
problems. 

We have shown that the Casimir-Polder interaction of two atoms is modified when the atoms are confined by perfectly conducting cylindrical shell, and their distance is comparable or larger than the radius of the shell.  This is due to the fact that Casimir interactions are not pairwise additive.  We have considered the limits of zero and high temperature, both in the retarded and non-retarded cases. In all situations, the interaction is substantially reduced compared to free space and decays exponentially over the scale of the radius of the shell.  Although we assumed a perfectly conducting cylindrical shell, we expect that our conclusion remain qualitatively unchanged for general dielectric materials.

Our findings are partially complementary to other studies of the fluctuation induced interactions between confined atoms or molecules.  Previous work for two atoms placed between two parallel perfect metal plates~\cite{2_atoms_in_PM_capacitor_zeroT} showed that at zero temperature the interaction energy is reduced from the free space case with an exponential decay in the non-retarded limit, and increased in the retarded case with a change from $h^{-7}$ to $L^{-2}h^{-5}$ when $h$ is much bigger than the distance $L$ between the plates.  For the same geometry, an exponential decay was also found at any finite temperature \cite{2_atoms_in_PM_capacitor_highT}.  An exponential reduction of the interaction between the atoms has been also observed when they are confined in a rectangular waveguide and their distance is large compared to the diameter of the waveguide \cite{Atoms_in_rectangular_cavity}. Contrary to that, a recent study of two atoms placed inside a transmission line consisting of two concentric 
metallic cylinders demonstrated a huge amplification of the interaction between the atoms due to the one-dimensional character of {\it propagating} fluctuations in this geometry \cite{Shahmoon+2014}. 

If confinement leads to reduction or amplification of inter-molecular fluctuation forces has important consequences for the (non-)additivity of these forces, and hence the electromagnetic response of confined gases. Therefore, it is important to understand the mechanism that determines the effect of confinement on the interaction. In general,  a massive fluctuation mode leads to an exponential decay of the associated interaction energy. In waveguides, massive modes are realized by evanescent modes due to the existence of finite cutoff frequencies for these modes. Clearly, all modes in our cylindrical shell, and also in the rectangular waveguide of Ref.~ \cite{Atoms_in_rectangular_cavity} are massive, explaining an exponential reduction of the interaction. However, for two parallel conducting plates, there exists a massless mode only for TM polarization \cite{Milonni}. Since this propagating mode contributes for two particles with electric polarizability only in the retarded limit, the interaction is not 
exponentially suppressed but enhanced in this limit. Hence the situation resembles that of the transmission line. It would be interesting to establish corresponding results for non-equilibrium effects in confined geometries.

\acknowledgments
The research leading to these results has received funding from the People
Programme (Marie Curie Actions) of the European Union's Seventh
Framework Programme (FP7/2007-2013) under REA grant agreement nº 302005.
P.~R.-L. acknowledge helpful discussions with A.~A. Saharian.
P.~R.-L.’s research has also been supported by Projects MOSAICO, UCM (Grant No. PR34/07-15859), MODELICO (Comunidad de Madrid), ENFASIS (Grant No.FIS2011-22644, Spanish Government) in the Departamento de F\'isica Aplicada I and GISC of the Facultad de Ciencias F\'isicas of the Universidad Complutense of Madrid, by "Ministerio de Econom\'ia y Competitividad" (Spain) Grant No. FIS2012-38866-C05-01 in the Departamento de Matem\'aticas and GISC of the Universidad Carlos III de Madrid, by the Engineering and Physical Sciences Research Council under EP/H049797/1 in the Department of Physics of the Loughborough University.

\appendix
\section{Matrix elements}
\label{Appendix}

Here we provide explicit expressions for the elements of the various matrices of the
scattering approach for imaginary wave numbers $k=i\kappa$.

\subsection{Conversion matrices}

The elements of the conversion matrices from spherical vector multipoles to cylindrical vector multipoles are \cite{Ehsan}
\begin{align}
D_{l m M,n k_{z}M} & = \phantom{-} C_{l m}\frac{\sqrt{\kappa^{2} + k_{z}^{2}}}{\kappa}{P_{l}^{m}}'\left(- i \frac{k_{z}}{\kappa}\right)\delta_{nm},\\
D_{l m E,n k_{z}M} & =  \phantom{-} C_{l m}\frac{\kappa}{\sqrt{\kappa^{2} + k_{z}^{2}}}i m P_{l}^{m}\left(- i \frac{k_{z}}{\kappa}\right)\delta_{nm},\\
D_{l m M,n k_{z}E} & =  - D_{l m E,n k_{z}M},\\
D_{l m E,n k_{z}E} & =  \phantom{-} D_{l m M,n k_{z}M},
\end{align}
where
\begin{equation}
C_{l m} = (-1)^{l - m}\frac{4\pi}{\sqrt{ l(l + 1) }}\sqrt{ \frac{2l + 1}{4\pi} \frac{(l - m)!}{(l + m)!}}.
\end{equation}

\subsection{T--matrix of a perfect metal sphere}

For a perfect sphere of radius $R$, the T--matrix elements are well known. In the spherical wave basis they are given  in terms of the Bessel functions $I_l$ and $K_l$ by~\cite{Rahi-Emig},
\begin{align}
\mathcal{T}_{S\,\,l m M,l' m' M} & = - \delta_{ll'}\delta_{mm'}\frac{\pi}{2}\frac{I_{l + \frac{1}{2}}(\kappa R)}{K_{l + \frac{1}{2}}(\kappa R)} \, ,\\
\mathcal{T}_{S\,\,l m N,l' m' N} & = - \delta_{ll'}\delta_{mm'}\frac{\pi}{2}\frac{l I_{l + \frac{1}{2}}(\kappa R) - \kappa R I_{l - \frac{1}{2}}(\kappa R)}{l K_{l+ \frac{1}{2}}(\kappa R) + \kappa R K_{l - \frac{1}{2}}(\kappa R)} \, ,
\end{align}
and the elements coupling unlike polarizations vanish.

\subsection{T--matrix of an atom}

The T--matrix of a two-state atom can be expressed in terms of its dipolar polarizability as a diagonal matrix, whose non--vanishing elements are
\begin{align}
\mathcal{T}_{S\,1,m,M;1,m',M} &= - \delta_{m,m'}\frac{2\kappa^{3}}{3}\alpha_{1}^{MM},\\
\mathcal{T}_{S\,1,m,N;1,m',N} &= - \delta_{m,m'}\frac{2\kappa^{3}}{3}\alpha_{1}^{EE},
\end{align}
where the electric and magnetic polarizabilities at imaginary frequency $\kappa$ are given by
\begin{align}
\label{eq:polarizability_two_state}
\alpha_{1}^{EE} &= \frac{\alpha_{E}}{1+ d_{10}^{2}\kappa^{2}},\\
\alpha_{1}^{MM} &= \frac{\alpha_{M}}{1 + d_{10}^{2}\kappa^{2}},
\end{align}
and $d_{10} = c/\omega_{10}$ with the transition frequency $\omega_{10}$.

\subsection{T--matrix of a perfect metal cylinder}

We need the T-matrix of a perfect metal cylinder for both exterior and interior scattering.
For the case of exterior scattering, the non-vanishing elements are given by
\begin{align}
\mathcal{T}_{C\,\,n k_{z} N,n' k'_{z} N} & = - \delta(k_{z} - k'_{z})\delta_{nn'}\frac{I_{n}(Rp)}{K_{n}(Rp)}\, ,\\
\mathcal{T}_{C\,\,n k_{z} M,n' k'_{z} M} & = - \delta(k_{z} - k'_{z})\delta_{nn'}\frac{I'_{n}(Rp)}{K'_{n}(Rp)} \, ,
\end{align}
where $p = \sqrt{k_{z}^{2} + \kappa^{2}}$ and $R$ is the radius of the cylinder. For the case of interior scattering the elements are obtained by inversion of the elements for the exterior case,
\begin{align}
\hspace{-2mm}\mathcal{T}_{C\,\,n k_{z} N,n' k'_{z} N} & = - \delta(k_{z} - k'_{z})\delta_{nn'}\frac{K_{n}(Rp)}{I_{n}(Rp)}\, , \\
\mathcal{T}_{C\,\,n k_{z} M,n' k'_{z} M} & = - \delta(k_{z} - k'_{z})\delta_{nn'}\frac{K'_{n}(Rp)}{I'_{n}(Rp)} \, .
\end{align}

\subsection{Cylindrical translation matrices}

The translation matrices that relate regular cylindrical waves to  regular cylindrical waves with respect to a displaced origin are diagonal in the polarization with the non-vanishing elements given by \cite{Rahi-Emig}
\begin{align}
\mathcal{V}_{SC\,\,nk_{z},n'k'_{z}} & =  (-1)^{n+n'}I_{n - n'}(d\sqrt{k_{z}^{2} + \kappa^{2}})e^{-i(n - n')\theta_{SC}}\nonumber\\
&  \times e^{-ik_{z}X_{SC,z}}\delta(k_{z} - k'_{z}), \\
\mathcal{V}^{\dagger}_{CS,\,\,nk_{z},n'k'_{z}} & =  (-1)^{n+n'}I_{n - n'}(d\sqrt{k_{z}^{2} + \kappa^{2}})e^{-i(n - n')\theta_{CS}}\nonumber\\
&  \times e^{ik_{z}X_{CS,z}}\delta(k_{z} - k'_{z}),
\end{align}
where $\theta_{CS} = \theta_{SC}\,\,(\text{mod }2\pi)$. The elements of the matrix relating regular to outgoing waves (relevant to exterior scattering between two objects $A_1$ and $A_2$) are given by
\begin{align}
\mathcal{U}_{A_1 A_2\, nk_{z},n'k'_{z}} & =  (-1)^{n'}K_{n - n'}(d\sqrt{k_{z}^{2} + \kappa^{2}})e^{-i(n - n')\theta_{A_1A_2}}\nonumber\\
& \times e^{-ik_{z}X_{A_1A_2,z}}\delta(k_{z} - k'_{z}).
\end{align}

\subsection{Spherical translation matrices}
It turns out that it is useful for the computations presented in the main text to express the translation matrices for spherical waves in terms of those for cylindrical waves, using the conversion matrices $D_{l\,m\,Q,n\,k_{z}\,P}$. The elements of the spherical translation matrix can then be written as
\begin{align}
\label{U_cyl_2_sph}
\mathcal{U}_{A_1A_2\,l\,m\,Q,\tilde{l}\tilde{m}\tilde{Q}} & = 
\sum_{n,n',P}\int_{-\infty}^{\infty}\frac{dk_{z}}{2\pi}
\int_{-\infty}^{\infty}\frac{dk'_{z}}{2\pi} D_{l\,m\,Q,n\,k_{z}\,P}\nonumber\\
& \times\mathcal{U}_{A_1A_2\,n\,k_{z},n'\,k'_{z}}
D^{\dagger}_{n'\,k'_{z}\,P,\tilde{l}\tilde{m}\tilde{Q}}
\frac{C^{c}_{P}}{C^{s}_{\tilde Q}}.
\end{align}
Taking into account that the cylindrical translation matrix is diagonal in $k_{z}$, we can carry out the integral over $k'_{z}$. Finally, the translation matrices in spherical waves can be found in \cite{Rahi-Emig}.


\begin{thebibliography}{20}



\bibitem{London} F.~London, Trans.~Faraday Soc. {\bf 33}, 8 (1937).

\bibitem{VdW int. electrica} H. B. G. Casimir and D. Polder, Phys. Rev. \textbf{73}, 360 (1948).

\bibitem{Serry} F. M. Serry, D. Walliser, and G. J. Maclay, Journal of Applied Physics \textbf{84}, 2501 (1998).

\bibitem{Klimchitskaya2009} G.L. Klimchitskaya, U. Mohideen, and V.M. Mostepanenko, Rev. Mod. Phys. {\bf 81}, 1827 (2009).

\bibitem{Casimir_review} D. Dalvit, P. Milonni, D. Roberts, and F. Rosa, eds., Lecture Notes in Physics, Casimir Physics, vol. 834 (Springer, 2011).


\bibitem{Derivative_Expansion_PFA_energy} G. Bimonte and T. Emig and R. L. Jaffe and M. Kardar, EPL \textbf{97} 50001 (2012).

\bibitem{edges_and_tips} M. F. Maghrebi, S. J. Rahi, T. Emig, N. Graham, R. L. Jaffe, and M. Kardar, PNAS \textbf{108}, 6867 (2011).

\bibitem{Orientation_dependence} T. Emig, N. Graham, R. L. Jaffe, and M. Kardar, Phys. Rev. A \textbf{79}, 054901 (2009).

\bibitem{Object_in_spherical_cavity} S. Zaheer, S. J. Rahi, T. Emig and R. L. Jaffe. Phys. Rev. A \textbf{82}, 052507 (2010).

\bibitem{cylinder_in_cylindrical_cavity} D. A. R. Dalvit, F. C. Lombardo, F. D. Mazzitelli, and R. Onofrio, Phys. Rev. A \textbf{74}, 020101(R) (2006).

\bibitem{EGJK}T. Emig, N. Graham, R.L. Jaffe and M. Kardar. Phys. Rev. Lett. \textbf{99}, 170403 (2007).

\bibitem{Rahi-Emig}S. J. Rahi, T. Emig, N. Graham, R. L. Jaffe, and M. Kardar. Phys. Rev. D \textbf{80}, 085021 (2009).

\bibitem{Multiscattering_Lambrecht}A. Lambrecht P. A. Maia Neto and S. Reynaud. \textit{New Journal of Physics} \textbf{8}, 243 (2006). 




\bibitem{Real_metallic_cylinders}E. Noruzifar, T. Emig, and R. Zandi, Phys. Rev. A \textbf{84}, 042501 (2011).
\bibitem{Real_Cylinder_and_plane}E. Noruzifar, T. Emig, U. Mohideen, and R. Zandi, Phys. Rev. B \textbf{86}, 115449 (2012).
\bibitem{Cylinder_and_plane}T. Emig, R. L. Jaffe, M. Kardar, and A. Scardicchio, Phys. Rev. Lett. \textbf{96}, 080403 (2006).
\bibitem{Cylinders_and_Plates}S. J. Rahi, T. Emig, R. L. Jaffe, and M. Kardar, Phys. Rev. A \textbf{78}, 012104 (2008).
\bibitem{spheres_and_plates} R. Zandi, T. Emig, and U. Mohideen, Phys. Rev. B \textbf{81}, 195423 (2010).
\bibitem{inclined_cylinders} P. Rodriguez-Lopez and T. Emig, Phys. Rev. A \textbf{85}, 032510 (2012).
\bibitem{Ehsan}E. Noruzifar, P. Rodriguez-Lopez, T. Emig and R. Zandi. Phys. Rev. A \textbf{87}, 042504 (2013).
\bibitem{Teo_Sphere_Cylinder}L. P. Teo. Phys. Rev. D \textbf{87}, 045021 (2013).

\bibitem{2_atoms_in_PM_capacitor_zeroT} J. Mahanty and B. W. Ninham. J. Phys. A: Math., Nucl. Gen. \textbf{6}, 1140 (1973).
\bibitem{2_atoms_in_PM_capacitor_highT} M. Bostr\"om, J. J. Longdell and B. W. Ninham. Phys. Rev. A \textbf{64}, 062702 (2001).

\bibitem{Ellingsen1}S. \AA{}. Ellingsen, S. Y. Buhmann and S. Scheel. Phys. Rev. A \textbf{82}, 032516 (2010).

\bibitem{PFA_Atom_Cylinder}V.B. Bezerra, E.R. Bezerra de Mello, G.L. Klimchitskaya, V.M. Mostepanenko, and A.A. Saharian. Eur. Phys. J. C \textbf{71}:1614 (2011).

\bibitem{Atoms_in_rectangular_cavity} E. Shahmoon and G. Kurizki. Phys. Rev. A \textbf{87}, 062105 (2013).

\bibitem{Shahmoon+2014} E.~Shahmoon, I.~Mazets, and G.~Kurizki, PNAS {\bf 111}, 10485 (2014).



\bibitem{Fosco_2012a} C. D. Fosco, F. C. Lombardo, and F. D. Mazzitelli, Phys. Rev. D {\bf 86}, 045021 (2012). 

\bibitem{SJohnson}S. J. Rahi, A. W. Rodriguez, T. Emig, R. L. Jaffe, S. G. Johnson and M. Kardar.  Phys. Rev. A \textbf{77}, 030101 (2008).
\bibitem{Emig_Pablo}P. Rodriguez-Lopez, S. J. Rahi and T. Emig. Phys. Rev. A \textbf{80}, 022519 (2009).

\bibitem{2_atoms_in_dielectric_capacitor} M. Marcovitch and H. Diamant.  Phys. Rev. Lett. \textbf{95}, 223203 (2005).


\bibitem{Milonni} P.~W.~Milonni, 
{\it The quantum vacuum : an introduction to quantum electrodynamics} (Academic Press, 1994)













\end{thebibliography}
\end{document}